\documentclass[twocolumn,superscriptaddress,prb]{revtex4}

\usepackage{graphicx}
\usepackage{epsfig}
\usepackage{verbatim}
\usepackage{bm}

\def\up{\uparrow}
\def\down{\downarrow}
\newcommand{\gammabar}{\ensuremath\gamma\kern-0.53em-}

\usepackage{times}
\usepackage{amsmath}
\usepackage{amsfonts}
\usepackage{amssymb}
\usepackage{graphicx}
\usepackage{pst-all}
\usepackage{leftidx}

\usepackage{tikz}

\makeatletter
\def\l@subsubsection#1#2{}
\makeatother

\begin{document}

\title{Particle-Hole Symmetry and the Composite Fermi Liquid}
\author{Maissam Barkeshli}
\affiliation{Station Q, Microsoft Research, Santa Barbara, California 93106-6105, USA}
\author{Michael Mulligan}
\affiliation{Department of Physics, Stanford University, Stanford, CA 94305, USA}
\affiliation{Kavli Institute for Theoretical Physics, University of California, Santa Barbara, California 93106, USA}
\author{Matthew P. A. Fisher}
\affiliation{Department of Physics, University of California, Santa Barbara, California 93106, USA}
\date{\today}

\begin{abstract}
The half-filled Landau level is widely believed to be described by the Halperin-Lee-Read theory of the composite Fermi liquid (CFL). 
In this paper, we develop a theory for the particle-hole conjugate of the CFL, the Anti-CFL, 
which we argue to be a distinct phase of matter as compared with the CFL. 
The Anti-CFL provides a possible explanation of a recent experiment [Kamburov et. al., Phys. Rev. Lett. 113, 196801 (2014)] 
demonstrating that the density of composite fermions in GaAs quantum wells corresponds to the electron density when the filling fraction $\nu < 1/2$ 
and to the hole density when $\nu > 1/2$. We introduce a local field theory for the CFL and Anti-CFL
in the presence of a boundary, which we use to study CFL - Insulator - CFL junctions, and the interface between the Anti-CFL and CFL. 
We show that the CFL - Anti-CFL interface allows partially fused boundary phases in which ``composite electrons'' 
can directly tunnel into ``composite holes,'' providing a non-trivial example of transmutation between topologically distinct quasiparticles. 
We discuss several observable consequences of the Anti-CFL, including a predicted resistivity jump at a first order transition 
between uniform CFL and Anti-CFL phases. We also present a theory of a continuous quantum phase transition between the CFL 
and Anti-CFL. We conclude that particle-hole symmetry requires a modified view of the half-filled Landau level, in the presence 
of strong electron-electron interactions and weak disorder, as a critical point between the CFL and the Anti-CFL.
\end{abstract}

\maketitle


\section{Introduction}

A powerful way of understanding the rich variety of fractional quantum Hall (FQH) states realized in two dimensional electron 
systems in the lowest Landau level is in terms of the theory of composite fermions.\cite{jain1989,jainCF} In the simplest 
version of this theory, the combination of the strong Coulomb interactions and the applied magnetic field make it favorable 
for electrons to bind to two units of flux quanta of an emergent gauge field, thus transforming into composite fermions. 
At filling fraction $\nu = 1/2$, the composite fermions see on average zero magnetic field and therefore form a
Fermi liquid-like state which may be referred to as a composite Fermi liquid (CFL).\cite{kalmeyer1992,halperin1993} 
Remarkably, the main sequence of incompressible FQH states can then be understood naturally in terms of integer 
quantum Hall (IQH) states of the composite Fermi liquid. Even-denominator incompressible FQH states, such as is observed at
$\nu = 5/2$,\cite{willett1987} are believed to result from pairing of the composite fermions.\cite{Moore1991,greiter1991,read2000}

The introduction of the CFL was a remarkable milestone in the study of the FQH effect, as the CFL explained experimentally 
observed transport anomalies at $\nu=1/2$ and the scaling of ratios of the energy gaps in nearby incompressible FQH states.
Subsequent experiments directly verified the existence of an emergent Fermi surface of composite fermions.\cite{willett1990,willett1993,du1993,kang1993,willett1997} 
It is therefore widely believed that the physics in the neighborhood of $\nu = 1/2$ in the lowest Landau level in GaAs 
quantum wells is well-described by the CFL. In addition, the CFL theory implies that strong interactions between the 
composite fermions and the fluctuating emergent gauge field results in a state of matter that is distinct from a conventional 
Landau Fermi liquid. Thus, the CFL provides a paradigmatic example of a metallic non-Fermi liquid that can be investigated 
both theoretically and experimentally.

In recent years,\cite{levin2007,lee2007antipf} it has been emphasized that in the absence of Landau level mixing, the Hamiltonian for the lowest Landau level 
has a strict particle-hole symmetry, associated with transforming $\nu \rightarrow 1-\nu$. \footnote{See Ref. \onlinecite{girvin1984} for 
an early study of particle-hole symmetry.}
At $\nu =1/2$ it is therefore equally natural to consider the state obtained by attaching two units of flux quanta 
to the holes of the filled $\nu=1$ Landau level instead of the electrons. If the CFL were particle-hole symmetric, such a 
construction in terms of flux attachment to holes would be an equivalent way of describing the CFL. However, for its 
paired descendant, the Moore-Read Pfaffian state,\cite{Moore1991} it was shown that particle-hole conjugation does yield a topologically distinct phase, 
which lead in particular to the prediction of the Anti-Pfaffian state.\cite{levin2007,lee2007antipf} This suggests that the particle-hole conjugate of
the CFL might also be a distinct state. 

In this paper, we develop a theory of the particle-hole conjugate of the CFL, which we refer to as the Anti-CFL. We provide arguments that the Anti-CFL 
is a distinct state of matter as compared with the original CFL, which therefore must break particle-hole symmetry. This provides a 
possible explanation of a recent experiment\cite{kamburov2014} in which the Fermi wave vector and therefore the density of composite 
fermions was carefully measured through the period of magnetoresistance oscillations in the presence of a one-dimensional periodic grating. 
Remarkably, it was found that the density of composite fermions corresponds to the electron density when $\nu < 1/2$, and to the density 
of holes when $\nu > 1/2$, thus providing evidence that the Anti-CFL is indeed realized when $\nu > 1/2$ in the LLL. This picture thus 
requires a modified view of the half-filled Landau level with strong interactions and 
negligible Landau level mixing as being described not by the CFL, as is the conventional understanding, but rather as a critical point between the CFL and Anti-CFL. 

\subsection{Particle-Hole Symmetry: General Considerations}
\label{overview}

Hints that the composite Fermi liquid at $\nu = 1/2$ might require particle-hole symmetry breaking has come from a 
number of previous works.\cite{kivelson1997,levin2007,lee2007antipf} In Ref. \onlinecite{kivelson1997}, it was pointed 
out that particle-hole symmetry requires that the electrical Hall conductivity satisfy $\sigma_{xy} = \frac{1}{2} \frac{e^2}{h}$, 
and that the only way to obtain such a Hall conductivity within the CFL mean-field theory (in the presence of disorder
which is statistically particle-hole symmetric) is to assume that the composite fermions have an extremely large 
Hall conductivity with the precise value, $\sigma_{xy}^{\text{cf}} = -\frac{1}{2} \frac{e^2}{h}$. 
This suggests a fundamental tension between particle-hole symmetry and the CFL state because the composite fermions 
should see zero effective field on average, so one instead expects $\sigma_{xy}^{\text{cf}} \approx 0$. 
In this paper, we suggest that a natural resolution of this tension is that the CFL state spontaneously breaks particle-hole 
symmetry (in the limit of zero Landau-level mixing) and one must also consider its particle-hole conjugate, the Anti-CFL, 
in order to properly describe the physics about $\nu=1/2$. These considerations should also be relevant to systems with 
weak Landau level mixing, as we will discuss. 

More recently, it has been noticed that the Moore-Read Pfaffian state,\cite{Moore1991} which is a candidate to explain 
the filling fraction $5/2$ plateau in GaAs systems,\cite{willett1987} breaks particle-hole symmetry. \cite{levin2007,lee2007antipf} 
Its particle-hole conjugate, referred to as the Anti-Pfaffian, has distinct topological properties and is, therefore, 
a different topological phase of matter. The Pfaffian can be thought of as a $p_x + i p_y$ paired state of the composite 
fermions in the CFL. This suggests that the parent CFL state might also break particle-hole symmetry. The particle-hole 
conjugate of the CFL, the Anti-CFL, would then be the parent state of the Anti-Pfaffian. 

If the CFL state spontaneously breaks particle-hole symmetry, then a clean system at filling fraction $\nu  = 1/2$
must lie at a phase transition between the CFL and Anti-CFL. In a clean (disorder-free) system, the nature of this transition will either be first order or continuous. 
In Sec. \ref{trans} we provide a theory of a continuous transition in the clean limit. In the presence of disorder that locally favors CFL or Anti-CFL, 
the transition will necessarily be continuous, as we discuss in more detail in Sec. \ref{trans}. A quantum critical point between the CFL and Anti-CFL 
would control a broad region of the finite temperature phase diagram in the vicinity of $\nu = 1/2$. 

Tuning away from half-filling explicitly breaks particle-hole symmetry and therefore favors one state over the other. 
There are thus two possibilities: Either the CFL is preferred for $\nu < 1/2$ and the Anti-CFL for $\nu > 1/2$, or vice versa.
The question of which one of these possibilities is realized depends on microscopic details of the interactions. 
Generally, one would expect that the CFL is more stable when $\nu < 1/2$, while the Anti-CFL is preferred for $\nu > 1/2$. 
This intuition comes from the energetics of model wave functions. Wave functions for FQH states in the lowest Landau level 
at fillings $\nu < 1/2$, such as the $1/3$ Laughlin wave function, are holomorphic in the complex coordinates of the 
electrons and may be interpreted as describing an integer Hall state of ``composite electrons'' (where the statistical flux is attached to the electrons) 
rather than an integer state of composite holes. 
Therefore, it is natural to expect, by continuity with the wave function of the Laughlin state, that the 
CFL state, which attaches flux to the electrons, would be more stable for $\nu < 1/2$. Analogous considerations for the 
holes at FQH filling fractions $\nu>1/2$ suggest that the Anti-CFL controls the physics when $\nu>1/2$. If the clean system 
for $\nu < 1/2$ is controlled by the CFL, we note that disorder may locally favor puddles of Anti-CFL, and vice versa for $\nu > 1/2$. 

\subsection{Summary of Results}

Due to the length of this paper, here we will briefly summarize some of the main results of our paper. 

\subsubsection{Bulk field theory}

The effective field theory describing fluctuations about the Anti-CFL state is different from that of the CFL state, and is presented in 
Eqs. (\ref{antiCFL1})-(\ref{antiCFL2}). Specifically, the CFL and Anti-CFL theories have a different set of emergent gauge fields 
coupled to the composite fermions. The structure of the action for these gauge fields suggest that the composite
fermions in the CFL represent fundamentally different nonlocal degrees of freedom of the electron fluid as compared with
the composite fermions of the Anti-CFL. Consequently, $p$-wave pairing of the composite fermions of the CFL leads to topologically
distinct phases of matter as compared with $p$-wave pairing the composite fermions of the Anti-CFL. To highlight the distinction between these two types of
composite fermions, we refer to the composite fermions of the CFL as \it composite electrons\rm, and the composite fermions of the Anti-CFL
as \it composite holes\rm. As expected, the robust topological features of
the incompressible FQH states away from half filling can be readily accessed 
in terms of both composite electrons or composite holes. 

\subsubsection{Boundary physics}

An important property of the CFL is that the single electron correlations, as probed for example by the 
local frequency-dependent tunneling density of states, decay exponentially in the bulk of the CFL,\cite{kim1994,he1993} 
but as a power-law on the edge. \cite{shytov1998,levitov2001} Therefore, although the system is gapless everywhere, the 
electron effectively has a finite correlation length in the bulk and infinite correlation length along the boundary. 
Here, we introduce a novel, local formulation of the CFL field theory in the presence of a boundary which accounts 
for the structure of these edge correlations in terms of a robust chiral scalar field along the edge (see Eqn. (\ref{CFLedge})).
We expect that this formulation will be useful more generally also for the study of boundary criticality in quantum phase transitions 
between incompressible FQH states. 

We apply our formulation of the CFL boundary theory to deduce the theory of the Anti-CFL in the presence of a boundary (see Eqn. (\ref{acflEdge})). 
We find that a distinction between the CFL and the Anti-CFL is in the structure of their respective 
boundary theories: for example, the edge of the Anti-CFL hosts an additional chiral field that is inherited from the filled Landau level.  

We study a variety of interfaces, including CFL - insulator (I) - CFL junctions, the interface between CFL and Anti-CFL, and
their paired descendants: Pfaffian - I - Pfaffian junctions, and Pfaffian - Anti-Pfaffian interfaces. 
Our study of the CFL - I - CFL junction reveals how to understand the healing together of two adjacent CFL states through electron
tunneling, leading to effectively a single statistical gauge field stretching across the interface and the elimination of the chiral 
boundary fields. We show that CFL - I - CFL junctions support in principle three distinct boundary phases, which are 
distinguished by whether composite fermions can directly tunnel across the junction and whether the electron correlations 
decay exponentially or algebraically along the interface. 

\begin{figure}
	\centering
	\includegraphics[width=3.2in]{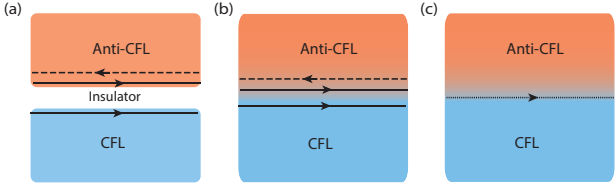}
	\caption{\label{cfliAcfl} (a) Interface between the CFL and Anti-CFL state, separated by a thin insulator. When electron tunneling across the interface is weak and irrelevant, the interface is in a decoupled phase. (b) Possibility of a partially fused phase along the interface. Composite electrons can directly tunnel into composite holes across the interface, but the chiral boson edge fields are unconstrained, leading to a protected power-law tunneling density of states along the interface. 
}
\end{figure}

\begin{figure}
	\centering
	\includegraphics[width=3.2in]{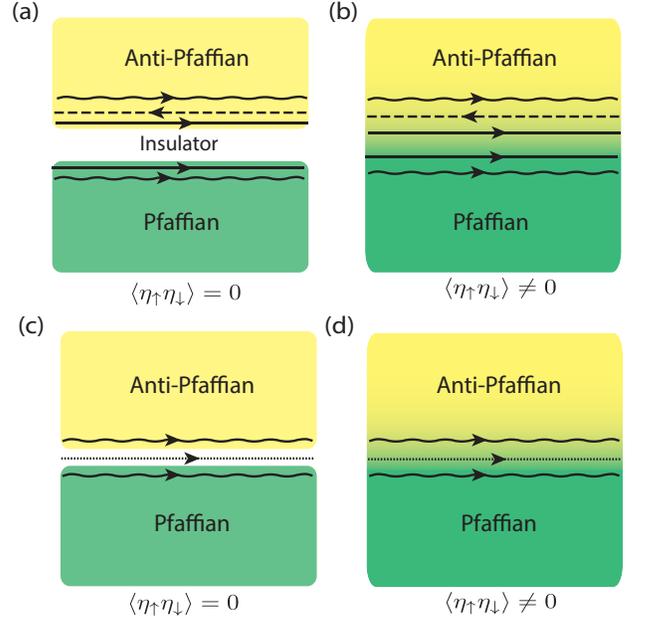}
	\caption{\label{PfApf} The interface between the Pfaffian and Anti-Pfaffian states hosts four topologically distinct interface phases. The wavy lines
depict the chiral Majorana fermion modes, while the solid lines depict the chiral $\nu=1/2$ Luttinger liquid modes, and the dashed line depicts the 
chiral $\nu=1$ Luttinger liquid mode. 
(a) The decoupled interface. (b) Partially fused interface 1, where $\langle \eta_\up \eta_\down \rangle \neq 0$, which allows the neutral fermion from
the Pfaffian to directly tunnel across the interface as the neutral fermion in the Anti-Pfaffian. 
(c) Partially fused interface 2, where the boson modes have been partially gapped, leaving behind a neutral $\nu=1$ chiral boson mode. 
$\langle \eta_\up \eta_\down \rangle = 0$, so that the neutral fermion from the Pfaffian cannot directly tunnel across the interface. 
(d) Partially fused interface 3, which is the same as (c), but $\langle \eta_\up \eta_\down \rangle = 0$, allowing the neutral fermion to directly
tunnel across the interface. }
\end{figure}

Our study of the interface between the CFL and Anti-CFL reveals several possible distinct interface phases (see Fig. \ref{cfliAcfl}). 
These include ``partially fused'' interfaces where composite electrons of the CFL can directly tunnel across the interface
as composite holes of the Anti-CFL. Since the composite electrons of the CFL and the composite holes of the 
Anti-CFL appear not to be related to each other by any local operators, this provides a non-trivial and experimentally testable 
example of transmutation between topologically distinct quasiparticles.\footnote{See Ref. \onlinecite{barkeshli2014sledge} 
for a different example studied recently in the context of quantum spin liquids.} 
As a consequence, our results also imply that the boundary between the Pfaffian and Anti-Pfaffian, while hosting 
topologically robust chiral edge modes, host a number of distinct boundary phases (see Fig. \ref{PfApf}), some of which allow 
the neutral fermion of the Pfaffian to tunnel directly into the neutral fermion of the Anti-Pfaffian. Importantly, we find that 
within our field theoretic formulation, the boundary between the CFL and Anti-CFL must contain chiral scalar fields
along the interface. The ``minimal'' interface phase we find contains a neutral chiral fermion field. 

\subsubsection{Continuous phase transition between CFL and Anti-CFL}

Within our field theoretic framework, we find that tuning from the CFL to the Anti-CFL state requires passing through a phase transition. 
In fact, as we describe in Sec. \ref{trans}, the field theory we develop for the Anti-CFL can be used to describe a continuous phase transition between the CFL and Anti-CFL in a clean system. 
The critical point can be described in terms of a neutral Dirac fermion, coupled to the various statistical gauge fields and 
composite fermion fields in the system (see Eqn. (\ref{transitionF})). The neutral Dirac fermion at the critical point can be naturally understood in terms of the theory of the CFL - Anti-CFL 
interface described above, which necessarily contains a chiral fermion field at the interface. 

In Sec. \ref{trans}, we also present a detailed analysis of the effect of gauge fluctuations on the critical theory, to show that it remains continuous
even beyond the mean-field limit. We further analyze the finite-temperature phase diagram and discuss the appearance of two crossover 
temperature scales, as has appeared in several other studies of quantum criticality in fractionalized systems with emergent quasiparticle 
Fermi surfaces \cite{senthil2008,barkeshli2012hlr}.

An interesting consequence of this study is that one can also understand the quantum phase transition between the Pfaffian and Anti-Pfaffian states
in terms of a massless, neutral Dirac fermion mode. This implies that if experiments were to realize both the Pfaffian and Anti-Pfaffian states by tuning across
half-filling (say in the second Landau level), then the gap to charged excitations need not close, and therefore electrical charge transport measurements alone
may not show any indication of a quantum phase transition. 

\subsubsection{Observable consequences}

We discuss two main observable signatures that demonstrate the distinction between the CFL and Anti-CFL states. 
First, we consider the conductivity of the system at $\nu = 1/2$, in the presence of disorder which is statistically particle-hole symmetric
(so that $\rho_{xx} \neq 0$). Using the Ioffe-Larkin sum rules\cite{IoffeLarkin89,halperin1993} for the resistivity of the system, we observe that the CFL state at $\nu = 1/2$ 
possesses a Hall conductivity (see Eqn. (\ref{cflsigmaxy}))
\begin{align}
\sigma_{xy}^{\text{CFL}} < \frac{1}{2} \frac{e^2}{h}, 
\end{align}
while the Hall conductivity of the Anti-CFL state is (see Eqn. (\ref{acflsigmaxy}))
\begin{align}
\sigma_{xy}^{\text{ACFL}} > \frac{1}{2} \frac{e^2}{h}.
\end{align}
Therefore, we find that at a first order transition between uniform CFL and Anti-CFL states, the system will exhibit a jump in
the Hall conductivity. The magnitude of the jump is set by the longitudinal resistivity $\rho_{xx}$, as explained in Sec. \ref{resistivitySec}.
We find similar jumps in both longitudinal and transverse components of the resistivity tensor. 
The Hall conductivities obey the ``sum rule" at half-filling:
\begin{align}
\sigma_{xy}^{{\rm CFL}} + \sigma_{xy}^{{\rm ACFL}} = {e^2 \over h}.
\end{align}
Within this theory, systems that do not exhibit such a jump
either lie at a continuous transition between these two states and do not realize uniform CFL or Anti-CFL states arbitrarily close 
to half-filling, or have strong Landau level mixing and thus explicitly break particle-hole symmetry even at $\nu = 1/2$. 
At the continuous transition between the CFL and Anti-CFL presented in Sec. \ref{trans}, the Hall conductivity is equal to ${1 \over 2} {e^2 \over h}$, within linear response, such that the sum rule continues to be obeyed.

A second observable consequence of the distinction between the CFL and Anti-CFL is as follows. 
The wave vector (and therefore the density) of composite fermions can be measured close to $\nu=1/2$ 
through, \it e.g.\rm, magnetoresistance oscillations in the presence of a periodic potential. 
As the system is tuned from the CFL to the Anti-CFL by tuning the filling fraction through $\nu = 1/2$, the composite fermion density
does not evolve smoothly, but instead possesses a kink as it transitions from being set by the electron density to being set by the hole density. 
This singularity in the evolution of the composite fermion density through $\nu=1/2$ 
further suggests that the CFL and Anti-CFL are distinct phases of matter. If, instead, the system remains in the CFL state on both sides of $\nu = 1/2$,
then the composite fermion density would have a smooth dependence on filling fraction. In Sec. \ref{expPrinceton}, we analyze in detail the 
experiment of Ref. \onlinecite{kamburov2014}, which performs such a measurement of the composite fermion density through magnetoresistance oscillations, 
and we show that it can be explained by the existence of the Anti-CFL state for $\nu > 1/2$. 

Moreover, in Sec. \ref{expTrans}, we suggest a further experiment to detect the coherent tunneling of composite electrons to composite holes across the
interface between CFL and Anti-CFL states, by having the periodic potential modulation be strong enough to force the system into alternating
strips of CFL and Anti-CFL. 

The rest of this paper is organized as follows. In Sec. \ref{review}, we provide a review of the conventional theory of the CFL and its connection to the main sequence of incompressible FQH states. In Sec. \ref{antiCFLbulk} we develop the bulk effective field theory of the Anti-CFL and describe its relation to the main sequence of incompressible FQH states. In Sec. \ref{edgeTheories} we describe the edge theories of the CFL and Anti-CFL, and use them in Sec. \ref{interfaces} to study a variety of interfaces, including CFL - I - CFL junctions and CFL - Anti-CFL junctions, Pfaffian-I-Pfaffian junctions, and Pfaffian-Anti-Pfaffian interfaces.  In Sec. \ref{trans} we develop a field theory of a direct transition between the CFL and Anti-CFL. In Sec. \ref{exp} we describe experimental consequences that can provide probes of the difference between the Anti-CFL and CFL, including an analysis of the experiment of Ref. \onlinecite{kamburov2014}. In Appendix \ref{dualitySec}, we develop a more general understanding of particle-hole conjugates of fermionic FQH states, in terms of vortex duals of bosonic FQH states. 

\section{Review of CFL}
\label{review}

In this section, we will review the theory of the CFL as introduced in Ref. \onlinecite{halperin1993}. 
We set the fundamental constants $e = \hbar = c = 1$, so that the flux quantum $\phi_0 = hc/e = 2\pi$. 

\subsection{Bulk effective field theory at $\nu = 1/2$}

To motivate the CFL description of the half-filled Landau level, we begin with the field theory for (spin-polarized) electrons:
\begin{align}
\label{elTheory}
\mathcal{L}_{\rm electron} = &c^\dagger (i \partial_t + A^E_{t} + \mu_e) c - \frac{1}{2m_e} c^\dagger (i \partial_i + A^E_{i})^2 c 
\nonumber \\
&+ \int d^2r' V(r-r') c^\dagger c (r) c^\dagger c(r'),
\end{align}
where $i = x,y$, $c$ is the electron operator, $A^{E}$ is the background or external electromagnetic field, 
which includes the constant background magnetic field, $\mu_e$ is the chemical potential, 
$m_e$ is the electron band mass, and $V(r - r')$ describes the leading electron-electron interactions.

The CFL theory\cite{halperin1993} starts by considering an equivalent theory to that in Eqn. (\ref{elTheory}) where $2$ units of (magnetic) flux quanta of an emergent statistical gauge field $a$ are attached to a ``composite fermion'' field $\psi$:
\begin{align}
\label{hlrAction}
\mathcal{L}_{\rm CFL} = &\frac{1}{2} \frac{1}{4\pi} a \partial a +\psi^\dagger (i \partial_t +a_t+ A^E_{t} + \mu_e) \psi 
\nonumber \\
&-\frac{1}{2m_\psi}\psi^\dagger (i \partial_i +a_i + A^E_{i})^2 \psi 
\nonumber \\
&+\int d^2r' V(r-r') \psi^\dagger \psi (r) \psi^\dagger \psi (r'),
\end{align}
where we have used the notation $a \partial a \equiv \epsilon_{\mu \nu \lambda} a_\mu \partial_\nu a_\lambda$, $\mu,\nu,\lambda = t,x,y$.
The original electron $c$ consists of a composite fermion $\psi$ bound to two units of flux of the $a$ gauge field. $m_\psi$ is the effective
mass of the composite fermions, which may in principle differ from the band mass of the electrons at long wavelengths due to interaction 
effects. 

At half-filling, when the electron filling fraction $\nu \equiv 2\pi n_e/B = 1/2$, there exists a mean-field solution to the equations of motion of Eqn. (\ref{hlrAction}) 
where the total flux of $a$ is equal and opposite to the total applied magnetic flux $B$ of $A^E$. The composite fermions then effectively see no magnetic field on average, and thus
form a Fermi sea with wave vector $k_F = \sqrt{2 m_e \mu_e}$. Importantly, the composite fermion density is equal to the electron density $n_e = k_F^2/4\pi$.
Fluctuations about this mean-field ground state are described by composite fermions $\psi$, in zero 
net effective magnetic field, coupled to the fluctuations of the $U(1)$ gauge field $a$ about its mean-field value. 
If the coupling between the composite fermions and the fluctuations of $a$ were set to zero, the composite fermions would form a Landau Fermi liquid; 
in the presence of such a coupling, however, the composite fermions acquire an anomalous self-energy and must instead be described by a non-trivial 
interacting fixed point that is believed to be of a non-Fermi liquid character. 
In order to ensure consistency with various general constraints, including certain sum rules and Kohn's theorem, sophisticated methods have been developed
to treat the gauge fluctuations.\cite{simon1998}

In Section \ref{antiCFLbulk}, we will develop a theory -- much like the one above -- for the holes of the filled Landau level. 
It is convenient to refer to the composite fermions of the CFL as {\it composite electrons} in order to distinguish them from the {\it composite holes} of the theory to be introduced in Sec. \ref{antiCFLbulk}. 

It is useful to note that the above theory can be obtained in a different way, through a parton construction. We write the electron operator as
\begin{align}
\label{partonCFL}
c = b \psi,
\end{align}
where $b$ is a charge-$e$ boson and $\psi$ is a neutral fermion with respect to the electromagnetic field. 
The decomposition in Eqn. (\ref{partonCFL}) results in a local $U(1)$ redundancy, 
$b \rightarrow e^{- i\Lambda(r)} b$, $\psi \rightarrow e^{i \Lambda(r)} \psi$, under which all physical operators must be invariant, and an emergent $U(1)$ gauge field $a$ transforming as $a \rightarrow a + \partial \Lambda$.
Next, we consider a mean-field ansatz where $b$ forms a bosonic $1/2$-Laughlin (incompressible) FQH state, 
$a$ is zero on average, and $\psi$ forms a Fermi sea. 
The effective field theory for such a mean-field ansatz is:
\begin{align}
\label{partonCFL2}
\mathcal{L}_{\rm CFL} =& -\frac{2}{4\pi} \tilde{a} \partial \tilde{a} + \frac{1}{2\pi} (a - A^E)\partial \tilde{a} 
\nonumber \\
&+\psi^\dagger (i \partial_t + a_t + \mu_e) \psi - \frac{1}{2m_\psi}\psi^\dagger (i \partial_i + a_i)^2 \psi + \cdots,
\end{align}
where the boson current is $j^b_\mu = \frac{1}{2\pi} \epsilon_{\mu \nu \lambda} \partial_\nu \tilde{a}_\lambda$, $m_\psi$ is the effective mass of $\psi$, 
$\mu_e$ is the electron chemical potential, and the $\cdots$ indicate possible higher order interactions.
The electromagnetic field $A^E$ represents deviations from the fixed background magnetic field at exactly
half-filling. Redefining $a \rightarrow a - A^E$, and integrating out $\tilde{a}$ yields Eqn. (\ref{hlrAction}). 
Note that in the long-wavelength description in Eqn. (\ref{partonCFL2}), the boson $b$ is created by inserting $2$ units of flux of
$a$, because $b$ forms a $1/2$ Laughlin state and has unit charge under the $a$ gauge field. Therefore the electron operator at long wavelengths becomes 
precisely $\psi$ bound to two units of flux of $a$, as expected. 
This presentation is useful because it allows one to consider the transition to nearby 
phases, such as to a conventional Landau Fermi liquid, under the application of an external periodic potential.\cite{barkeshli2012hlr}

\subsection{Single-Particle Properties}

The CFL is a metallic state: it is compressible and possesses a 
finite longitudinal resistivity in the presence of disorder. 
However, electron tunneling into the bulk of the CFL is exponentially suppressed at low energies.
Assuming a (Fourier transformed) electron-electron interaction of the form, $V(|q|) \sim 1/|q|^\eta$ with 
$0 \leq \eta \leq 2$, which physically corresponds to a three-dimensional Coulomb interaction evaluated
in the two-dimensional system, the electron tunneling density of states $A(\omega)$ decays as,\cite{kim1994,he1993}
\begin{align}
\label{bulkTDOS}
A(\omega) \propto e^{- (\omega_0/\omega)^{1/(2-\eta)}},
\end{align}
where $\omega_0 = \omega_0(\eta)$ is a finite constant that depends upon the nature of the interaction.
$\eta = 0$ corresponds to short-ranged interactions, while $\eta = 1$ corresponds to (un-screened) Coulomb interactions.

The composite electrons display a sharp Fermi surface, 
however, they are not well defined quasiparticles (for $\eta \leq 1$) due to their coupling to the fluctuating emergent gauge field.\cite{halperin1993, nayakwilczekNFL1, nayakwilczekNFL2} 
Within an RPA treatment of this interaction, the composite electrons obtain a self-energy correction,
\begin{align}
\Sigma^{(\eta)}_\psi(\omega) \propto i |\omega|^{\frac{2}{3-\eta}},
\end{align}
in the regime where the composite electron's frequency is less than its momentum.
This self-energy correction implies a vanishing quasiparticle weight for $\eta \leq 1$. 
The self-energy also directly implies the following finite temperature corrections to scaling of the specific heat.\cite{holsteinnortonpincus,halperin1993}
For short-ranged interactions $(\eta = 0)$, the specific heat scales as
\begin{align}
C \propto T^{2/3} .
\end{align}
For un-screened Coulomb interactions ($\eta = 1$), the specific heat instead receives a contribution
\begin{align}
C \propto T \ln (k_F/T) , 
\end{align}
and the self-energy of the composite electrons takes the marginal Fermi liquid form,\cite{MFL}
\begin{align}
\Sigma^{(\eta =1)}_{\psi}(\omega) \propto \omega \ln(k_F/\omega).
\end{align}

\subsection{CFL Wave Function}

A wave function for the CFL in the half-filled Landau level was previously presented,\cite{rezayi1994} and is given by:
\begin{align}
\label{cflwfn}
\Psi_{CFL}(\{r_i\}) = \mathcal{P}_{LLL} \prod_{i < j} (z_i - z_j)^2 \psi_{FS}(\{r_i\}) e^{- \sum_i |z_i|^2/4l_B^2},
\end{align}
where $\mathcal{P}_{LLL}$ is the projection onto the lowest Landau level, $z_i = r_{i,x} + i r_{i,y}$ are the complex 2D coordinates of the $i$th electron, 
$\psi_{FS}$ is a Slater determinant wave function describing free fermions with a Fermi surface, and $l_B = B^{-1/2}$ is the magnetic length. 

\subsection{Relation to the main sequence of incompressible FQH states}
\label{fqhHLR}

One of the major successes of the CFL theory is that it can describe a wide series of incompressible FQH states seen experimentally in terms of integer QH states of the composite electrons. 
As we move away from half-filling, the composite electrons feel an effective magnetic field,
\begin{align}
B_{\rm eff} = B - 4 \pi n_e.
\end{align}
Therefore, the composite electrons are at an effective filling fraction,
\begin{align}
\nu_{\text{eff}}^{-1}= \nu^{-1} - 2.
\end{align}
When $\nu_{\text{eff}} = p$, where $p$ is an integer, so that $\nu = \frac{p}{2p+1}$, the composite electrons fill $p$ Landau levels.
At low energies, we may integrate out the composite electrons $\psi$ to obtain an effective theory:
\begin{align}
\mathcal{L}_p^{\text{CFL}} = \frac{1}{2} \frac{1}{4\pi} a \partial a + \frac{p}{4\pi} (a + A^E) \partial (a + A^E),
\end{align}
which is equivalent to
\begin{align}
\label{CFLeffective}
\mathcal{L}_p^{\text{CFL}}= 
-\frac{2}{4\pi} \tilde{a} \partial \tilde{a} + \frac{1}{2\pi} a \partial \tilde{a} + \frac{p}{4\pi} (a + A^E) \partial (a + A^E),
\end{align}
as can be verified by integrating out $\tilde{a}$. 
The topological order of this theory is therefore encoded in a $K$-matrix,\footnote{Recall that the $K$-matrix is defined
in terms of an effective Chern-Simons theory as $\mathcal{L} = K_{IJ} a^I \partial a^J$.\cite{wen04} }
\begin{align}
\label{CFLKmatrix}
K = \left(\begin{matrix} -2 &1 \\ 1 & p \end{matrix} \right),
\end{align}
which provides a formulation of the effective field theory of the $\nu = \frac{p}{2p+1}$ hierarchy states.\cite{lopezfradkin91,lopez1999}
The central charge of the edge theory can be obtained by noticing a $A^E \partial A^E$ term for the background 
electromagnetic field in Eqn. (\ref{CFLeffective}) with {\it non-zero} coefficient proportional $p$. 
This indicates the presence of $p$ filled Landau levels whose chiral edge modes combined with those implied by the $K$-matrix in Eqn. (\ref{CFLKmatrix}) give a chiral central charge -- a quantity that coincides, for Abelian states, with the number of left-moving edge modes minus the number of right-moving edge modes, equal to $p - {\rm sign}(p) + 1$.

A remarkable prediction of the CFL theory, which is borne out by experiments,\cite{du1993} is the dependence of the ratio of energy gaps of the resulting FQH states on the 
composite electron Landau level index. Specifically, the CFL theory predicts that when the composite electrons fill $p$ Landau levels, 
the energy gap of the resulting state,
\begin{align}
E_\nu^{\text{cfl}} = \frac{e B_{\rm eff}}{m_\psi} = \frac{e}{m_\psi} \frac{2\pi n_e}{|p|},
\end{align}
where $n_e$ is the electron density, $p^{-1} = \nu^{-1} - 2$, $m_\psi$ is the renormalized composite electron mass, and we have restored the electron charge $e$. 
Therefore, in terms of the (electron) filling fraction,
\begin{align}
E_\nu^{\text{cfl}} = {2\pi e n_e \over m_\psi} |\nu^{-1} - 2|. 
\end{align}

\subsection{Paired composite electron states}
\label{pairedCFLSec}

It is well-known that the Moore-Read Pfaffian state can be understood as a state where the composite electrons have formed
a $p_x + i p_y$ paired state. In general, one can consider any odd angular momentum $l$ pairing of the composite electrons, with a corresponding
chiral central charge $l/2$ in the boundary state of the composite electrons. Including the chiral charged mode, the chiral central charge
of the resulting FQH state is $c = 1 + l/2$. The topological properties of the resulting FQH state can be readily obtained. The system
has four topologically distinct Abelian quasiparticles: those corresponding to local (topologically trivial) excitations, a neutral 
fermion associated with the Bogoliubov quasiparticle of the composite electron paired state, and charge $\pm e/2$ Laughlin 
quasiparticles. In addition, the $\pm \pi$ vortices of the composite electron state carry charge 
$\pm e/4$. The $e/2$ Laughlin quasiparticle is associated with $2\pi$ units of external flux, and therefore has statistics $\pi/2$ 
The $\pm \pi$ vortices have statistics $\theta_{\pm} = 2\pi (1/16 + l/16 )$, which is a sum of contributions
of $\pi$ flux in the charge sector and the contribution from the composite electron sector, which forms one of the Ising states 
in Kitaev's 16-fold way.\cite{kitaev2006} These results are summarized in Table \ref{pairedCFLtable} .

\begin{table}
\begin{tabular}{ c | c  }
Charge (modulo $e$) & Statistics, $\theta$ (modulo $\pi$) \\
\hline
  $0$ & $0$ \\
  $0$ & $0$  \\
  $e/2$ & $\pi/2$  \\
  $-e/2$ & $\pi/2$  \\
  $e/4$ & $\pi (l+1)/8$ \\
  $-e/4$ & $\pi (l+1)/8$ \\
\end{tabular}
\caption{\label{pairedCFLtable} Charges and statistics for the topologically non-trivial quasiparticles of the paired composite electron states with angular momentum 
$l$ and chiral central charge $c = 1 + l/2$. }
\end{table}

\section{Theory of Anti-CFL }
\label{antiCFLbulk}

\subsection{Bulk Field Theory}

Here we develop a bulk effective field theory for the particle-hole conjugate of the CFL, which we refer to as the Anti-CFL. 
In this case, the statistical flux is attached to the holes of a filled Landau level, rather than to the 
electrons. 

\subsubsection{Derivation of hole theory}

In order to provide a description of the Anti-CFL, we first need to 
derive an effective theory for the holes in the lowest Landau level.\cite{lee2007antipf}
We start with the field theory in terms of the electron fields in Eqn. (\ref{elTheory}) and attach one unit of flux to each electron, in
the direction opposite to the external field. This transmutes their statistics, yielding composite bosons\cite{zhang1989} at an effective
filling fraction equal to $1$ within a mean-field approximation. The effective action for such a theory is:
\begin{align}
\mathcal{L}_{\tilde{b}} = &\tilde{b}^* (i\partial_t + a_t + A^E_{t}) \tilde{b} - \frac{1}{2m_{\tilde{b}}} 
|(i \partial_i + a_i + A^E_{i}) \tilde{b}|^2  
\nonumber \\
&+V(|\tilde{b}|)  +\frac{1}{4\pi} a \partial a,
\end{align}
where $\tilde{b}$ is a complex scalar field representing the composite boson, $a$ is a $U(1)$ gauge field, $V(|\tilde{b}|)$ is an interaction potential that encodes boson density-density interactions 
and $m_{\tilde{b}}$ is the boson mass which may differ from $m_e$ upon performing the duality. 
Next, we perform boson-vortex duality:\cite{dasgupta1981,fisher1989}
\begin{align}
\label{vortex1}
\mathcal{L}_{\tilde{b}_v} =& \tilde{b}_v^* (i \partial_t + \tilde{a}_t) \tilde{b}_v - \frac{1}{2m_v} |(i \partial_i + \tilde{a}_i) \tilde{b}_v|^2
\nonumber \\
&+ V(|\tilde{b}_v|)  - \frac{1}{2\pi} \tilde{a} \partial (a + A^E) + \frac{1}{4\pi} a \partial a ,
\end{align}
where $\tilde{b}_v$ represents the dual vortex field of $\tilde{b}$, 
$m_v$ is the vortex effective mass, and the $U(1)$ gauge field $\tilde{a}$ represents the conserved particle current of $\tilde{b}$,
\begin{align}
\label{bCurrent}
j^{(\tilde{b})}_{\mu} = \frac{1}{2\pi} \epsilon_{\mu \nu \lambda} \partial_\nu \tilde{a}_\lambda.
\end{align} 
Shifting $a \rightarrow a + A^E$, integrating out $a$, and subsequently shifting $\tilde{a} \rightarrow \tilde{a} + A^E$ gives
\begin{align}
\mathcal{L}_{\tilde{b}_v} = & \tilde{b}_v^* (i \partial_t + \tilde{a}_t - A^E) \tilde{b}_v - \frac{1}{2m_v} |(i \partial_i + \tilde{a}_i - A_E) \tilde{b}_v|^2
\nonumber \\
&+ V(|\tilde{b}_v|)  - \frac{1}{4\pi} \tilde{a} \partial \tilde{a} + \frac{1}{4\pi} A^E \partial A^E . 
\end{align}
The CS term for $\tilde{a}$ attaches one unit of flux to $\tilde{b}_v$, thereby transmuting it into a fermion $h$. Therefore, at long-wavelengths, 
the above theory is equivalent to:
\begin{align}
\label{holeTheory}
\mathcal{L}_{\rm hole} =& h^\dagger (i \partial_t - A^E_{t}  + \mu_h) h - \frac{1}{2m_h} h^\dagger (i \partial_i - A^E_{i} )^2 h
\nonumber \\
&+ \frac{1}{4\pi} A^E \partial A^E + \cdots,
\end{align}
where the $\cdots$ include higher order gradient and interaction terms, $h$ represents holes of a filled lowest Landau level (LLL), and we have allowed the effective mass of the holes $m_h$ to be renormalized relative to $m_v$ upon assuming the duality between bosons attached to one unit of flux and holes attached to zero flux in the presence of a background magnetic field. $\mu_h$ is the chemical potential of the holes and is given in terms of the hole density $n_h = 2 m_h \mu_h/4\pi$.

The Lagrangian in Eqn. (\ref{holeTheory}) is the particle-hole conjugate with respect to the filled lowest Landau level of the electron Lagrangian in Eqn. (\ref{elTheory}).
As expected, Eq. (\ref{holeTheory}) shows that the hole $h$ carries charge opposite to that of the electron with respect to the background electromagnetic field $A^E$.
If the hole density is depleted to zero, the system exhibits an integral quantum Hall effect with Hall conductivity $\sigma_{xy} = 2\pi$. In other words,
the vacuum state of the holes is correctly described by a filled Landau level. 

From an operator point of view, the above sequence of transformations can be understood as follows. 
As before, the electron $c$ can be written as
\begin{align}
\label{parton}
c = \tilde{b} f,
\end{align}
where $\tilde{b}$ is a boson and $f$ is a fermion field. These fields may be taken to carry opposite charge under an emergent $U(1)$ gauge field $a$.
We consider a mean-field ansatz where $f$ forms a $\nu_f = 1$ IQH state. Let us suppose that the field $\tilde{b}_v$ describes the vortices 
of the boson field $\tilde{b}$. Then, we can consider the fermion operator:
\begin{align}
h = f^\dagger \tilde{b}_v .
\end{align}
We can verify that $h$ is in fact a gauge-invariant, physical operator, and describes the holes. To see this, let us write down an effective theory in
terms of $\tilde{b}$. We first introduce a gauge field $a^f$ to describe the conserved current of the $f$ fermions: 
$j^{(f)}_{\mu} = \frac{1}{2\pi} \epsilon_{\mu \nu \lambda} \partial_\nu a^f_\lambda$. The effective action then takes the form
\begin{align}
\mathcal{L}_{\tilde{b}} = &\tilde{b}^* (i\partial_t + a_t + A^E_{t}) \tilde{b} - \frac{1}{2m_{\tilde{b}}} |(i \partial_i + a_i + A^E_{i}) \tilde{b}|^2  
\nonumber \\
&+V(|\tilde{b}|)  -\frac{1}{4\pi} a^f \partial a^f + \frac{1}{2\pi} a\partial a^f . 
\end{align}
Now let us consider the vortex dual of $\tilde{b}$:
\begin{align}
\mathcal{L}_{\tilde{b}_v} = & \tilde{b}_v^* (i \partial_t + \tilde{a}_t) \tilde{b}_v + \frac{1}{2m_v} |(i \partial_i + \tilde{a}_i) \tilde{b}_v|^2 + V(|\tilde{b}_v|)  
\nonumber \\
&-\frac{1}{4\pi} a^f \partial a^f + \frac{1}{2\pi} (a + A^E) \partial a^f + \frac{1}{2\pi} a \partial \tilde{a} ,
\end{align}
where, as before, $\tilde{b}_v$ represents the vortices of the field $\tilde{b}$, and the current of $\tilde{b}$ is given in terms of $\tilde{a}$ through eq. (\ref{bCurrent}). 
The above theory shows that $\tilde{b}_v$ is charged under the gauge field $\tilde{a}$. The Chern-Simons terms indicate that 
a unit charge of $\tilde{a}$ is attached to $-1$ unit of flux of $a$ and $-1$ unit of flux of $a^f$. Minus one unit of 
flux of $a^f$, in turn, corresponds to $-1$ unit of the $f$ fermion. Therefore, the operator $h = f^\dagger \tilde{b}_v$ is a 
gauge-invariant, physical operator in the low energy effective theory, and can be seen to correspond to the hole in the LLL. 

\subsubsection{Anti-CFL Lagrangian}

Now that we have obtained the Lagrangian describing holes in the Landau level, 
it is straightforward to obtain the Anti-CFL theory. 
We simply take Eqn. (\ref{holeTheory}), and attach two flux quanta to the hole field $h$, obtaining:
\begin{align}
\label{antiCFL1}
\mathcal{L}_{\text{ACFL}} = &-\frac{1}{2} \frac{1}{4\pi} a \partial a + \chi^\dagger (i \partial_t + a_t - A^E_{t} + \mu_h) \chi
\nonumber \\
&- \frac{1}{2m_h} \chi^\dagger (i \partial_i + a_i - A^E_{i})^2 \chi
+ \frac{1}{4\pi} A^E \partial A^E + \cdots.
\end{align}
Here, $\chi$ is a fermion field describing the ``composite hole,'' which is the analog in the Anti-CFL of the composite electron $\psi$ in the CFL. 
The $\cdots$ represent higher-order interactions. As usual, the sign of the Chern-Simons term is chosen so that the 
flux of $a$ is attached in such a way so as to cancel out the background applied magnetic field, in a flux-smearing mean-field approximation. 

There is an important relation between the electron density $n_e$, hole density $n_h$, and background magnetic field $B$:
\begin{align}
\label{elholerelation}
n_e + n_h = {B \over 2\pi},
\end{align}
within the first Landau level. We will use this relation to express physical quantities symmetrically with respect to the electrons and holes.
Eqn. (\ref{elholerelation}) can be rewritten in another convenient form as
\begin{align}
n_h = (\nu^{-1} - 1) n_e,
\end{align}
where $\nu = 2\pi n_e/B$ is the filling fraction. 

The effective field seen by the composite holes is
\begin{align}
B_{\text{eff}} = - B + 4 \pi n_h.
\end{align}
The first term arises because the composite holes couple to the external field with the opposite sign as compared to 
electrons. The second term is due to the fact that each composite hole is attached to $+2$ units of flux, in contrast to the composite electrons which are attached to $-2$ units of flux. 
Using the relation in Eqn. (\ref{elholerelation}),
we find:
\begin{align}
\label{holeeffB}
B_{\text{eff}} = B - 4 \pi n_e.
\end{align}
Interestingly, the composite holes feel an {\it identical} effective magnetic field as the composite electrons.
In particular, this implies that a composite hole of the Anti-CFL moving with velocity $v$ in a magnetic field $B_{\rm eff}$ 
feels a Lorentz force perpendicular to its motion which is identical to the Lorentz force that would have been 
felt by a composite electron of the CFL, at the same magnetic field and with the same velocity. In other words,
the trajectory of the composite holes in the Anti-CFL and composite electrons in the CFL are the same, for a given velocity, magnetic field, and electron density.

An alternative, useful presentation of the theory in Eqn. (\ref{antiCFL1}) is given by:
\begin{align}
\label{antiCFL2}
\mathcal{L}_{\text{ACFL}} = & \frac{2}{4\pi} \tilde{a} \partial \tilde{a} + \frac{1}{2\pi} a \partial \tilde{a} + 
\chi^\dagger (i \partial_t + a_t - A^E_{t}) \chi
\nonumber \\
&- \frac{1}{2m_h} \chi^\dagger (i \partial_i + a_i - A^E_{i})^2 \chi
\nonumber \\
&- \frac{1}{4\pi} A \partial A + \frac{1}{4\pi} A^E \partial A ,
\end{align}
where $\tilde{a}$ and $A$ are both emergent dynamical gauge fields. 
Integrating out $\tilde{a}$ and $A$ in Eqn. (\ref{antiCFL2}) gives Eqn. (\ref{antiCFL1}). Eqn. (\ref{antiCFL2}) is manifestly well-defined on 
closed surfaces, in contrast to Eqn. (\ref{antiCFL1}), because the coefficient of the resulting CS terms are integral. 
Moreover, in Sec. \ref{edgeTheories} we will find Eqn. (\ref{antiCFL2}) to be a more useful starting point
to describe the system in the presence of a boundary. 
The fluctuations of $A$ describe the dynamics of the filled lowest Landau level, which is the ``vacuum'' state of the holes. 

We emphasize that the composite hole, $\chi$, is topologically distinct from the composite electron, $\psi$. From Eqn. (\ref{antiCFL2}), we see that 
the composite hole $\chi$ corresponds to the hole $h$, attached to two units of statistical flux of a $U(1)_2$ CS gauge field $\tilde{a}$. 
On the other hand, the composite electron $\psi$ corresponds to the electron $c$, combined with two units of statistical 
flux of a $U(1)_{-2}$ CS gauge field (see Eqn. (\ref{partonCFL2})). A direct consequence of this is that a $p_x + i p_y$ paired state of the $\psi$ fermions,
which yields the Moore-Read Pfaffian state, is topologically distinct from a $p_x - i p_y$ paired state of the $\chi$ fermions, which yields the Anti-Pfaffian state
 (and even distinct from a $p_x + ip_y$ paired state of $\chi$). This suggests that \it $\psi$ and $\chi$ are not related to each other by any 
local operator; in other words, they represent topologically distinct degrees of freedom of the electron fluid. \rm In the CFL, $\psi$ represents 
the appropriate low energy degrees of freedom, while in the Anti-CFL, $\chi$ represents the appropriate low energy degrees of freedom by which to describe the state.

From the perspective of the parton construction, it is clear that the Anti-CFL state can be understood through Eqn. (\ref{parton}) by considering a mean-field ansatz 
where $f$ forms a $\nu_f = 1$ IQH state, while the vortices of $\tilde{b}$ form a $\nu_{v} = -1$ CFL state of bosons. In Appendix \ref{dualitySec}, we revisit 
this point in more detail, and show that, in general, particle-hole conjugates of fermionic FQH states can be understood in terms of vortex duals of bosonic FQH states. 
Vortex duality for bosonic states thus proves to be intimately related to particle-hole conjugation for fermionic states. 

\subsection{Relation to the main sequence of incompressible FQH states}

In the same way that the main series of FQH states can be understood as integer quantum Hall states of composite electrons of the CFL, those same states 
can also be understood as integer quantum Hall states of composite holes from the Anti-CFL. 

To see this, consider moving away from half-filling. Using Eqns. (\ref{holeeffB}) and (\ref{elholerelation}), we see that the composite holes are at an effective filling,
\begin{align}
\label{holefillingrelation}
(\nu^h_{\text{eff}})^{-1} = 2 - (1-\nu)^{-1} .
\end{align}
Therefore, if the composite holes are at filling $\nu^h_{\text{eff}} = -p$, for integer $p$, then the filling fraction of the electrons is $\nu = \frac{p+1}{2p+1} = 1 - \frac{p}{2p+1}$. 

Furthermore, we find that the resulting topological order is equivalent to the particle-hole conjugate of the state where the composite electrons of the CFL are at $\nu_{\text{eff}} = p$. 
This is implied from the above formula for the filling fraction. To verify this, let us analyze the topological order of the state where the composite holes $\chi$
are at $\nu^h_{\text{eff}} = -p$. Integrating out $\chi$ in eq. (\ref{antiCFL2}) under the assumption that $\chi$ is in such an IQH state then yields the effective theory,
\begin{align}
\mathcal{L}_{-p}^{\text{ACFL}} =&\frac{2}{4\pi} \tilde{a} \partial \tilde{a} + \frac{1}{2\pi} a \partial \tilde{a} - \frac{p}{4\pi} a \partial a \cr 
+ & \frac{p}{2\pi} A^E \partial (a + A) - \frac{1}{4\pi} A \partial A \cr 
- & \frac{p}{4\pi} A^E \partial A^E. 
\end{align}
This theory can be summarized by a $K$-matrix:
\begin{align}
K = - \left( \begin{matrix} 
1 & 0 & 0  \\
0 & -2 & -1 \\
0 & -1 & p \\
\end{matrix} \right) ,
\end{align}
which, by comparison with the results of Sec. \ref{fqhHLR}, is indeed the particle-hole conjugate of the state where the composite electrons of the CFL form a $\nu_{\text{eff}} = p$ IQH state. This shows that the incompressible FQH states at a given filling fraction $\nu$ that result from the CFL state are equally-well understood to arise from condensation of the composite holes of the Anti-CFL state. The chiral central charge is again accounted for by the $p$ filled Landau levels implied by the $-p A^E \partial A^E$ term.

Moreover, the ratio of the energy gaps of the resulting FQH states are the same as well. 
This follows from an estimate of the gap in terms of the effective magnetic field that the composite holes and composite electrons feel.
More explicitly, according to Eqn. (\ref{holefillingrelation}), a FQH state at electron filling $\nu = p/(2p+1)$ requires the composite holes to fill $p+1$ Landau levels.
Restoring the electromagnetic charge $e$, the resulting energy gap is estimated using:
\begin{align}
E_\nu^{\text{acfl}} = {e B_{\rm eff} \over m_\chi} = {- eB + 4 \pi  en_h \over m_\chi} =
\frac{e}{m_\chi} 2\pi  n_e |\nu^{-1} - 2|,
\end{align}
where $m_\chi$ is the renormalized composite hole mass.
Remarkably, we see that
\begin{align}
E_\nu^{\text{acfl}} = E_\nu^{\text{cfl}} \frac{m_\psi}{m_\chi},
\end{align}
which shows that the predicted energy gaps of the incompressible FQH states are the same within the CFL or 
Anti-CFL construction, as long as the composite electrons and composite holes have the same renormalized masses. 

\subsection{Paired composite hole states}
\label{pairedAntiCFLSec}

Let us now also consider paired states of the composite holes. The particle-hole conjugate of the Moore-Read Pfaffian,
the Anti-Pfaffian, corresponds to $p_x - i p_y$ pairing of the composite holes. More generally, one can consider 
odd angular momentum $l$ pairing of the composite holes, with a corresponding
chiral central charge $l/2$ in the boundary state of the composite holes. The chiral central charge of the 
boundary theory of the corresponding FQH state, after considering the charged modes and the background filled LL
is therefore $c = -1 +1 + l/2 = l/2$. The topological properties of the quasiparticles can be understood as follows.
There are four Abelian quasiparticles, corresponding to local (topologically trivial) excitations, a neutral fermion 
corresponding to the Bogoliubov quasiparticle of the paired composite hole state, and charge $\pm e/2$ Laughlin quasiparticles. 
In addition to these, there are the $\pm \pi$ vortices, with charge $\pm e/4$. These have fractional statistics
$\theta_{\pm} = 2\pi (-1/16 + l/16)$. The $-1/16$ contribution is the effect of the $\pi$ flux on the 
charged sector, which is reversed in chirality relative to the composite electron case considered in Sec. \ref{pairedCFLSec}, while the 
$l/16$ contribution is from the composite hole sector, which forms one of the Ising states in Kitaev's 16-fold way.\cite{kitaev2006}
These are summarized in Table \ref{pairedAntiCFLtable}.

\begin{table}
\begin{tabular}{ c |c  }
Charge (modulo $e$) & Statistics, $\theta$ (modulo $\pi$) \\
\hline
  $0$ & $0$ \\
  $0$ & $0$  \\
  $e/2$ & $\pi/2$  \\
  $-e/2$ & $\pi/2$  \\
  $e/4$ & $\pi (l-1)/8$ \\
  $-e/4$ & $\pi (l-1)/8$ \\
\end{tabular}
\caption{\label{pairedAntiCFLtable} Charges and statistics for the topologically non-trivial quasiparticles of the paired composite hole states with angular momentum 
$l$ and chiral central charge $c = l/2$. }
\end{table}

Interestingly, observe that there is a direct correspondence between the paired states obtained from the CFL theory, in Sec \ref{pairedCFLSec}, and those
obtained from the Anti-CFL. In particular, Tables \ref{pairedCFLtable}, \ref{pairedAntiCFLtable} show that the 
FQH state obtained when composite electrons of the CFL form an angular momentum $l$ paired state is topologically equivalent 
to the state obtained when the composite holes of the Anti-CFL form an angular momentum $l - 2$ state. 
In particular, this confirms a recent observation in Ref. \onlinecite{Soncompositefermion}, based on considerations of the shift\cite{wen1992} and central charge, 
that (1) the case where composite electrons of the CFL form the $l = -3$ paired state 
corresponds to the Anti-Pfaffian, and (2) the case where composite electrons form the $l = -1$ paired state gives rise to a 
state whose topological properties are particle-hole symmetric. 

\subsection{Wave function}

Since the Anti-CFL is a state where the holes have formed a CFL state at opposite magnetic field, its many-body wave function can be immediately written in terms of the many-body wave function of the CFL. It is given by:
\begin{align}
\label{acflwfn}
\Psi_{ACFL} (\{r_i\}) = \int \prod_{\alpha=1}^{N_h} d^2 \bar{r}_\alpha \Psi_{h}(\{r_i, \bar{r}_\beta\}) \Psi_{CFL}^*(\{\bar{r}_\beta \}) ,
\end{align}
where 
\begin{align}
\Psi_{h}(\{r_i, \bar{r}_i\}) =& \prod_{\alpha < \beta} (\eta_\alpha - \eta_\beta) \prod_{a,i} (\eta_\alpha - z_i) \prod_{i < j} (z_i - z_j) 
\nonumber \\
&\times e^{- \sum_i |z_i|^2/4l_B^2} . 
\end{align}
$r$, $\bar{r}$ are the real-space coordinates of the electrons and holes, respectively, 
while $z = r_x + i r_y$ and $\eta = \bar{r}_x + i \bar{r}_y$ are their complex coordinates. 
The indices take values: $\alpha, \beta = 1,\cdots, N_h$ and $i,j = 1,\cdots N_e$, where $N_h$ 
and $N_e$ are the number of electrons and holes, respectively. $\Psi_{h}(\{r_i, \bar{r}_i\})$ 
can be understood as a wave function of $N_h$ holes at fixed positions $\{\bar{r}_i\}$, 
in a $\nu  =1$ IQH state of $N_e$ electrons. The prefactor $\prod_{\alpha < \beta} (\eta_\alpha - \eta_\beta)$ is included to
ensure the proper normalization and Fermi statistics of the holes; without this factor, the integral in
eq. (\ref{acflwfn}) would vanish because $\Psi_{CFL}(\{\bar{r}_\beta \})$ is anti-symmetric in its coordinates. 

The wave function presented in eq. (\ref{acflwfn}) is technically difficult to work with. We may consider a potentially more tractable wave function by using the basis of lowest Landau level orbitals on a sphere instead of the real-space coordinates. To this end, consider a quantum Hall system on a sphere, with $N_\Phi$ flux quanta piercing the sphere. The lowest Landau level has $N_\Phi + 1$ single-particle orbitals, which we can label $l_i$, for $i = 0 , \cdots, N_\Phi$. In this basis, we can write the Anti-CFL wave function as
\begin{align}
\Psi_{ACFL}( \{l_i\}) = \sum_{\{ \bar{l}_\alpha \}} \Psi_h( \{l_i ,\bar{l}_\alpha \})   \Psi_{CFL}^*(\{ \bar{l}_\alpha\}) .
\end{align}
$\Psi_{CFL}(\{ l_\alpha \})$ is the CFL wave function, written in the orbital basis of the lowest Landau level. $\Psi_h$ is the wave function where the $i$th electron occupies orbital $l_i$ and the $\alpha$th hole occupies the orbital $\bar{l}_\alpha$. 

We note that Ref. \onlinecite{rezayi2000} studied the overlap of the CFL wave function (\ref{cflwfn}) with its particle-hole conjugate, and found a high (but not unit) overlap for up to $N = 16$ particles. In other words, the CFL wave function, while almost particle-hole symmetric for small system sizes, is not exactly particle-hole symmetric; we therefore expect that in the thermodynamic limit, the overlap of the CFL wave function with its particle-hole conjugate will indeed vanish. 

\section{Boundary theories of CFL and Anti-CFL}
\label{edgeTheories}

An important aspect of the CFL state is its behavior near any boundary of the system. Just like incompressible FQH states, the compressible CFL states exhibit qualitatively new features at their edge, as compared with the bulk. Specifically, while the electron tunneling density of states decays exponentially in the bulk as the frequency $\omega \rightarrow 0$, the edge tunneling density of states decays as a power law due to the existence of protected gapless edge fields. Tunneling into the edge of a CFL would therefore yield a power-law non-linear current-voltage characteristic, $I \propto V^\beta$, for some exponent $\beta > 0$. 

In this section, we introduce a theory for the CFL and Anti-CFL states in the presence of a boundary. We will show that in order to be gauge-invariant, the bulk effective field theories discussed in the previous sections must include additional chiral scalar fields at the boundary.\cite{wengaplessboundary, stoneedgewaves} 
We note that the theory we develop below, which describes both the bulk and boundary fields simultaneously in a local way, complements previous work. Previous studies\cite{shytov1998,levitov2001} of the boundary of the CFL developed the boundary theory by effectively integrating out the fermions in the bulk in the presence of strong disorder. This leads to non-analytic terms in the action, and precludes a direct description of both the low energy composite fermion and chiral boson fields near the boundary. We expect that the new formulation of this boundary theory may also be useful for studying more generally the boundaries of Chern-Simons theories with gapless bulk matter fields, such as describe quantum critical points in FQH systems. 

\begin{figure}
	\centering
	\includegraphics[width=3.2in]{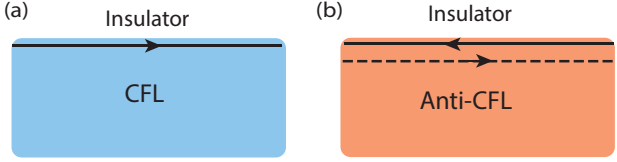}
	\caption{\label{edges} (a) The theory of the boundary of the CFL with a
          trivial insulator (vacuum) contains a chiral bosonic field at
          the edge. This bosonic field is analogous to the chiral edge mode of a bosonic $1/2$ Laughlin state. 
(b) The boundary of the Anti-CFL with a trivial insulator contains two fields. The left-moving field is similar
to that of a $1/2$ Laughlin edge, while the right-moving field is similar to a $\nu = 1$ IQH edge. To distinguish them,
the right-moving fields have been depicted with dashed lines. Consequently the fields
cannot be localized through backscattering and are topologically protected. 
}
\end{figure}

\subsection{Boundary theory for the CFL}
\label{boundaryCFL}

Let us begin with the bulk effective action for the CFL which we assume to be placed in a lower-half-plane geometry with boundary at $y=0$:
\begin{align}
\mathcal{L}_{\text{CFL}} =& -\frac{2}{4\pi} \tilde{a} \partial \tilde{a} + \frac{1}{2\pi} \tilde{a} \partial a
+\psi^\dagger (i \partial_t + a_t + A^E_{t} + \mu_e) \psi 
\nonumber \\
&- \frac{1}{2m_\psi}\psi^\dagger (i \partial_i + a_i + A^E_{i})^2 \psi + \cdots,
\end{align}
where we denote the composite electron mass by $m_\psi$.
Let us fix the gauge $\tilde{a}_t = 0$. In doing so, we must take into account the constraint $\frac{\delta \mathcal{L}}{\delta \tilde{a}_t} = 0$, which implies,
\begin{align}
\frac{2}{2\pi} \epsilon_{ij} \partial_i \tilde{a}_j = \frac{1}{2\pi} \epsilon_{ij} \partial_i a_j . 
\end{align}
This constraint can be solved by setting:
\begin{align}
\label{atilde}
\tilde{a}_i = \frac{1}{2} a_i + \partial_i \phi,
\end{align}
where $\phi$ is a real scalar field. 
Quantization of the flux of $\tilde{a}$ requires that $\phi$ be equivalent to $\phi+2\pi$: 
\begin{align}
\phi \sim \phi + 2\pi. 
\end{align}
Inserting $\tilde{a}$ in Eqn. (\ref{atilde}) back into the effective action, we obtain:
\begin{align}
S_{\text{CFL}} = \int_{y < 0} d^2x dt \mathcal{L}^{\text{bulk}}_{\rm CFL} + \int_{y = 0} dx dt \mathcal{L}^{\text{edge}}_{\rm CFL},  
\end{align}
where
\begin{align}
\label{CFLbulk}
\mathcal{L}^{\text{bulk}}_{\rm CFL} = &\frac{1}{2} \frac{1}{4\pi} a \partial a +\psi^\dagger (i \partial_t +a_t+ A^E_{t} + \mu_e) \psi 
\nonumber \\
&-\frac{1}{2m_\psi}\psi^\dagger (i \partial_i +a_i + A^E_{i})^2 \psi  + \cdots
\end{align}
\begin{align}
\label{CFLedge}
\mathcal{L}^{\text{edge}}_{\rm CFL} = &\frac{2}{4\pi} [(\partial_t \phi + a_t) \partial_x \phi + 
\frac{1}{4} a_t a_x - v (\partial_x \phi + \frac{1}{2} a_x)^2]
\nonumber \\
& - u (\partial_x \phi + \frac{1}{2} a_x) \psi^\dagger \psi.
\end{align}
The resulting action is invariant under the gauge transformations:
\begin{align}
\phi &\rightarrow \phi - \frac{1}{2} \Lambda,
\nonumber \\
a &\rightarrow a + \partial \Lambda
\nonumber \\
\psi &\rightarrow e^{i \Lambda} \psi,
\end{align}
where the real function $\Lambda$ is the gauge parameter.
The term involving $u$ above represents a density-density interaction between the composite electron and edge boson fields, while 
$v$ characterizes the velocity of the chiral $\phi$ excitations. We have included the terms involving $u$ and $v$ in the action
by hand, as they are not precluded by symmetry. 
One could also include a current-density coupling between the fermions and bosons, of 
the form $\psi^\dagger (i \partial_x + a_x) \psi (\partial_x \phi + \frac{1}{2} a_x)$, although this term is higher order and more irrelevant than the other
terms in the action. 

The electron creation operator on the edge is the gauge invariant operator
\begin{align}
c^\dagger(x,y = 0,t) \sim \psi^\dagger e^{-2 i \phi} . 
\end{align}
In terms of the parton construction of (\ref{partonCFL}), we see that $b = e^{2i \phi}$, consistent with the interpretation that $b$ forms a $1/2$ Laughlin state and $\phi$ represents the scalar field of the chiral Luttinger liquid edge of such a Laughlin state. 

Electron correlation functions along the boundary can be immediately computed within a mean-field approximation which ignores gauge fluctuations:
\begin{align}
\label{mfCorr}
\langle c^\dagger(x,t) c(0,0) \rangle \approx \langle \psi^\dagger(x,t) \psi(0,0)\rangle \langle e^{-2i (\phi(x,t) - \phi(0,0))} \rangle.
\end{align}
We see that single electron correlations decay algebraically because both the chiral field $\phi$ and the restriction to the boundary of the composite electron field are gapless. 
A detailed study of the quantitative effect of gauge fluctuations of $a$ on the edge correlation 
functions will be left for future work. 

\subsection{Boundary theory for the Anti-CFL}
\label{boundaryACFL}

The above method for deriving the edge theory of the CFL can be readily applied to derive the boundary theory for the Anti-CFL. As before, we start with the bulk theory in Eqn. (\ref{antiCFL2}). Going through essentially the same argument as in the previous section, we find the following effective action,
\begin{align}
S_{\text{ACFL}} = \int_{y < 0} d^2x dt ~ \mathcal{L}^{\rm bulk}_{\rm ACFL} + \int_{y = 0} dx dt ~ \mathcal{L}^{\rm edge}_{\rm ACFL},  
\end{align}
where
\begin{align}
\mathcal{L}^{\text{bulk}}_{\rm ACFL} = &-\frac{1}{2} \frac{1}{4\pi} a \partial a + \chi^\dagger (i \partial_t + a_t - A^E_{t} + \mu_h) \chi
\nonumber \\
&- \frac{1}{2m_\chi} \chi^\dagger (i \partial_i + a_i - A^E_i)^2 \chi,
\end{align}
and
\begin{widetext}
\begin{align}
\label{acflEdge}
\mathcal{L}^{\text{edge}}_{\rm ACFL} = 
&- \frac{2}{4\pi} [(\partial_t \phi_1 - a_t) \partial_x \phi_1 + 
\frac{1}{4} a_t a_x + v_1 (\partial_x \phi_1 - \frac{1}{2} a_x)^2]
+ \frac{1}{4\pi} \partial_x \phi_2(\partial_t \phi_2 - v_2 \partial_x \phi_2) - v_{12} (\partial_x \phi_1 - {1 \over 2} a_x) \partial_x \phi_2
\nonumber \\
&- u_1 (\partial_x \phi_1 - {1 \over 2} a_x) \chi^\dagger \chi - u_2 (\partial_x \phi_2)  \chi^\dagger \chi + [\xi(x) \chi e^{-2 i \phi_1 - i \phi_2} + {\rm h.c.}]
\end{align}
\end{widetext}
This action is invariant under the gauge transformations:
\begin{align}
\phi_1 &\rightarrow \phi_1 + \frac{1}{2} \Lambda,
\nonumber \\
a &\rightarrow a + \partial \Lambda
\nonumber \\
\chi &\rightarrow e^{i \Lambda} \chi,
\end{align}
where we again denote the gauge parameter by $\Lambda$.
The $\phi_2$ chiral edge field is a result of the filled Landau level and also satisfies: $\phi_2 \sim \phi_2 + 2 \pi$.

The $\phi_1$ and $\phi_2$ fields are counter-propagating. There are two local operators in the long wavelength effective field theory which can be identified as 
electron creation operators:
\begin{align}
c^\dagger_1 \sim \chi e^{-2 i \phi_1}, \;\; c^\dagger_2 \sim e^{i \phi_2}. 
\end{align}
In principle, the edge theory can also include electron tunneling between these two types of edge fields, mediated by the random coupling $\xi(x)$, as modeled by the last term in $\mathcal{L}^{\rm edge}_{\rm ACFL}$.

Importantly, in the mean-field limit, from the time-derivative of the $\phi_1$ term, we see that it can be interpreted as the chiral edge field of a bosonic $\nu = 1/2$ Laughlin FQH state,
while $\phi_2$ can be interpreted as a chiral edge mode of a $\nu = 1$ IQH of fermions. It is impossible for any type of backscattering to localize
these counter-propagating chiral fields.

\section{Theory of CFL and Anti-CFL Interfaces}
\label{interfaces}

A striking consequence of the existence of the Anti-CFL state is the nature of its interface with the CFL state. 
In order to understand the physical properties of this interface, it useful to first develop a theory of a CFL-insulator-CFL junction. 

\subsection{CFL - I - CFL Junctions}

\begin{figure}
	\centering
	\includegraphics[width=3.2in]{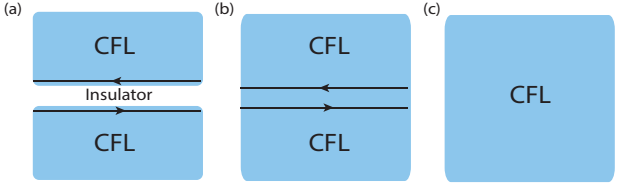}
	\caption{\label{cflicfl}
Illustration of three distinct possible interface phases at the CFL - I - CFL interface. (a) Decoupled interface, (b) partially fused interface, where composite fermions can propagate through and the electron has algebraic correlations along the interface, and (c) fully fused interface, where the system effectively forms a single CFL spanning across the interface, so that the electron has exponentially decaying correlations everywhere. 
}
\end{figure}

In this section, we develop an understanding of a junction between two CFL states, separated by a thin insulating barrier. We will see that there are three distinct possible interface phases:
\begin{enumerate}
\item \it Decoupled interface. \rm Electron tunneling between edges is weak and irrelevant. Composite electrons cannot tunnel across the boundary. Tunneling density of states decays as a power law at low energies at the interface, but exponentially in the bulk. 

\item \it Partially fused interface. \rm Electron tunneling across the interface is strong and/or relevant. Composite electrons can directly tunnel across the interface, but the counter-propagating bosonic edge fields remain gapless. Electron tunneling density of states decays as a power law at low energies at the interface, but exponentially in the bulk. 

\item \it Fully fused interface. \rm  Electron tunneling across the interface is strong and/or relevant, and the system has healed itself into effectively a single CFL across the interface. Electron tunneling density of states decays exponentially everywhere, and composite electrons can propagate through the interface. 
\end{enumerate}

Let us consider the insulating barrier to have a width $\epsilon$, with its center position at $y = 0$. The effective theory for this system can be written as
\begin{align}
\label{interfacefull}
S_{\rm CFL-I-CFL} = 
S_{\text{bulk}} + S_{\text{edge}},
\end{align}
where $S_{\text{bulk}} = S_{\text{bulk};\uparrow} + S_{\text{bulk};\downarrow}$ and $S_{\text{bulk}\uparrow} = \int_{y > \epsilon/2} d^2x dt \mathcal{L}_{\text{bulk};\uparrow} $ 
is the bulk action for the CFL on the upper-half plane, and $S_{\text{bulk};\downarrow}$, defined similarly, is the bulk action for the CFL on the lower-half plane. 
The bulk Lagrangian densities are
\begin{align}
\label{interfacebulk}
\mathcal{L}_{\text{bulk};\alpha} =& \frac{1}{2} \frac{1}{4\pi} a^\alpha \partial a^\alpha +\psi^\dagger_\alpha (i \partial_t +a^\alpha_{t}+ A^E_{t} + \mu_e) \psi_\alpha 
\nonumber \\
&-\frac{1}{2m_\psi}\psi^\dagger_\alpha (i \partial_i +a^\alpha_{i} + A^E_{i})^2 \psi_\alpha + \cdots,
\end{align}
where $\alpha = \uparrow, \downarrow$. 

Similarly, $S_{\text{edge}} = S_{\text{edge}; \uparrow} + S_{\text{edge};\downarrow} + S_{\text{edge};\uparrow\downarrow}$. 
$S_{\text{edge};\alpha} = \int_{y = \pm \epsilon/2} dx dt \mathcal{L}_{\text{edge};\alpha}$ is the edge action for the upper 
or lower boundary, which lies at $y = \pm \epsilon/2$, with
\begin{align}
\label{interfaceedge}
\mathcal{L}_{\text{edge}; \alpha} = & - \frac{2 \alpha}{4\pi} \Big[ (\partial_t \phi_{\alpha} - \alpha  a^{\alpha}_t) \partial_x \phi_{\alpha} \cr 
& + \alpha v_{\alpha} (\partial_x \phi_{\alpha} - \alpha {1 \over 2} a^{\alpha}_x)^2 + {1 \over 4} a^{\alpha}_t a^{\alpha}_x \Big]  \\
& - u_{\alpha} (\partial_x \phi_{\alpha} - \alpha {1 \over 2} a^{\alpha}_x) \psi_{\alpha}^\dagger \psi_{\alpha},
\end{align}
where $\alpha = \up/\down = \pm$.
The second equality is used when $\alpha$ appears as a coefficient, rather than a label.
Finally, $S_{\text{edge};\uparrow\downarrow} = \int dx dt ~ \mathcal{L}_{\text{edge};\uparrow \downarrow}$, with
\begin{align}
\label{interEdgeCFL}
\mathcal{L}_{\text{edge};\uparrow\downarrow} &=  \mathcal{L}_{\text{tun}; \uparrow\downarrow} 
+ v_{rs}^{\uparrow \downarrow} n_{r,\uparrow} n_{s, \downarrow} ,
\nonumber \\
\mathcal{L}_{\text{tun}; \uparrow\downarrow} &= t \psi_\uparrow^\dagger \psi_\downarrow e^{2i (\phi_\downarrow + \phi_\uparrow)} + {\rm h.c.},
\end{align}
where $r,s = \phi,\psi$, $n_{\phi, \alpha} = (\partial_x \phi_{\alpha} - \alpha {1 \over 2} a^{\alpha}_x)$, $n_{\psi, \alpha} = \psi^\dagger_\alpha \psi_\alpha$ are the boson and fermion densities, while
$v_{rs}^{\uparrow \downarrow}$ parameterizes the density-density interactions between excitations on each side of the interface. 
$\mathcal{L}_{\text{tun}; \uparrow\downarrow}$
describes electron tunneling between the upper and lower edges.

In order to understand the effect of the electron tunneling term, $\mathcal{L}_{\text{tun}; \uparrow\downarrow}$, let us rewrite it as follows:
\begin{align}
\label{eqTun1}
\mathcal{L}_{\text{tun}; \uparrow\downarrow} =& [ ( \psi_\up^\dagger \psi_\down + e^{- 2i  (\phi_\down + \phi_\up) - i\theta}) ( e^{2i (\phi_\down + \phi_\up) + i\theta} + \psi_\down^\dagger \psi_\up) 
\nonumber \\
&-1 - \psi_\up^\dagger \psi_\up \psi_\down \psi_\down^\dagger] |t|,
\end{align}
where we have written $t = |t| e^{i\theta}$. Note that for a clean system, $\theta = \epsilon x/l_B^2$, where $l_B = B^{-1/2}$ is the magnetic length. 
Introducing a complex Hubbard-Stratonovich field $\Phi_{\uparrow\downarrow}$, we can replace $\mathcal{L}_{\text{tun}; \uparrow\downarrow}$ by
\begin{align}
\label{edgequadratic}
\mathcal{L}_{\text{tun};\uparrow\downarrow} &= \sqrt{\frac{|t|}{2}} \left( \Phi_{\uparrow\downarrow} \psi_\up^\dagger \psi_\down + \Phi_{\uparrow\downarrow}^* e^{2i (\phi_\down + \phi_\up)+i\theta} + {\rm h.c.}\right) 
\nonumber \\
&-\frac{1}{2}|\Phi_{\uparrow\downarrow}|^2 + |t| \psi_\up^\dagger \psi_\up \psi_\down \psi_\down^\dagger,
\end{align}
where we have dropped the overall constant. Here, we have made use of the identity
\begin{align}
e^{\alpha \mathcal{O}_1 \mathcal{O}_2} = \int \mathcal{D} \Phi e^{-|\Phi|^2/2 + \sqrt{\frac{\alpha}{2}} (\Phi \mathcal{O}_1 + \Phi^* \mathcal{O}_2) } .
\end{align}
We can recover the original action by substituting into Eqn. (\ref{edgequadratic}) the solution to the equation of motion for $\Phi_{\up\down}$:
\begin{align}
\label{HSsolution}
\Phi_{\up \down} = \sqrt{2 |t|} \Big( e^{2 i (\phi_\down + \phi_\up) + i\theta} + \psi^\dagger_\down \psi_\up  \Big).
\end{align}
The last term in Eqn. (\ref{edgequadratic}) can be absorbed into the density-density couplings, and will be dropped from consideration below. 

Note that $\Phi_{\uparrow\downarrow}$ and the other fields transform under the gauge symmetry:
\begin{align}
\psi_\alpha &\rightarrow e^{i \Lambda_\alpha} \psi_\alpha,
\nonumber \\
a^{\alpha} &\rightarrow a^{\alpha} + \partial \Lambda_\alpha,
\nonumber \\
\Phi_{\up\down} &\rightarrow e^{i \Lambda_\up -i \Lambda_\down} \Phi_{\uparrow\downarrow} ,
\nonumber \\
\phi_{\up} &\rightarrow \phi_{\up} + \Lambda_{\up}/2, \cr
\phi_{\down} &\rightarrow \phi_{\down} - \Lambda_{\down}/2 ,
\end{align}
where the real functions $\Lambda_\alpha$ parameterize independent gauge transformations above and below the insulating barrier. 

We can obtain a self-consistent mean-field solution of the action in which gauge fluctuations are set to zero by replacing $\Phi_{\up\down}$ by its expectation value and minimizing the resulting action subject
to the constraint:
\begin{align}
\langle \Phi_{\up\down} \rangle = \sqrt{2 |t|} \langle \Big( e^{2 i (\phi_\down + \phi_\up) + i\theta} + \psi^\dagger_\down \psi_\up  \Big) \rangle.
\end{align}
This yields a mean-field solution 
which we denote as $\overline{\Phi}_{\up\down}$. 
\footnote{Equivalently, we may perform a saddle-point analysis by integrating out the composite electrons, gauge fields, and edge fields in order to obtain a gauge-invariant effective potential $V(|\Phi_{\up \down}|)$ for the Hubbard-Stratonovich field $\Phi_{\up \down}$.
The shape of the effective potential depends upon the effective parameters entering the action such as $v_{\up/\down}, u_{\up/\down}, v^{\up \down}_{r s}$ for $r,s = \phi, \psi$, and $t$ along with the coupling between the composite electrons and the statistical gauge fields.
We anticipate two general classes of (homogeneous) solutions that minimize the resulting effective potential to be identified by whether $\langle \Phi^*_{\up \down} \Phi_{\up \down} \rangle = 0$ or $\langle \Phi^*_{\up \down} \Phi_{\up \down} \rangle \neq 0$.}

In order to incorporate quantum fluctuations in the field $\Phi_{\up\down}$ about this mean-field solution, we write:
\begin{align}
\Phi_{\up\down} = \overline{\Phi}_{\up\down}e^{i a_{\uparrow\downarrow}}. 
\end{align}
Note that the amplitude fluctuations of $\Phi_{\up\down}$ about the mean-field value $\overline{\Phi}_{\up\down}$ are gapped.
Therefore, to describe physics at low energies, we can focus on the phase fluctuations of $\Phi_{\up \down}$, which are parameterized by $a_{\up\down}$.  

At low energies, the couplings in the edge Lagrangian will be renormalized due to fluctuations, and all possible terms consistent with
symmetries will be generated. In particular, gauge-invariant kinetic terms for $a_{\up\down}$ will be generated in the edge effective action 
that we anticipate to take the form:
\begin{align}
\label{secondsub}
\mathcal{L}_{\text{tun};\up\down} = &\kappa \overline{\Phi}_{\up\down}^2 (a^{\up} - a^{\down} - \partial a_{\up\down})^2 +
\overline{\Phi}_{\up\down} \Big( t_1 e^{i a_{\up\down}} \psi_\up^\dagger \psi_\down
\nonumber \\
& + {\rm h.c}. + t_2 \cos( 2 \phi_\up + 2\phi_\down - a_{\up\down}) \Big).
\end{align}
The first term above comes with a factor of $\overline{\Phi}_{\up\down}^2$ as it arises from a kinetic term for $\Phi_{\up \down}$ 
of the form $|(i \partial + a_\up - a_\down)\Phi_{\up \down}|^2$ upon the expansion of $\Phi_{\up\down}$ about $\overline{\Phi}_{\up\down}$. 
$\kappa \neq 0$ is a phenomenological parameter, $t_1$ and $t_2$ are renormalized parameters for terms that were already present
in Eqn. (\ref{edgequadratic}). We have also assumed for simplicity that there is no additional constant phase in the cosine term above. 

Notice that $a_{\up\down}$ can be interpreted as an emergent gauge field that bridges the upper and lower CFLs. 
The term  $\int_{y=0} dt dx (a^{\up} - a^{\down} - \partial a_{\up\down})^2$ plays the role of a Maxwell term across the 
boundary. 

We see that when $\overline{\Phi}_{\up \down} \neq 0$, composite electrons can directly tunnel across the boundary, due to the gauge-invariant composite electron hopping term $\psi_\up^\dagger \psi_\down e^{ia_{\up\down}}$.
Therefore, when $\overline{\Phi}_{\up\down} \neq 0$, we have effectively a single $U(1)$ gauge field $a$ defined everywhere, with $a_{\up\down}$ being equal to the line integral of $a$ across the boundary. 

Note that if $e^{i 2(\phi_\up + \phi_\down) - i a_{\up\down}}$ is irrelevant, the bosons will be remain gapless at low energies. If instead $e^{i 2(\phi_\up + \phi_\down) - i a_{\up\down}}$ is relevant, then $2(\phi_\up + \phi_\down) - a_{\up \down}$ will be pinned and become massive.

We thus see the appearance of three possible phases:

(1) $\overline{\Phi}_{\up\down} = 0$. Here, the bosons remain gapless as the coefficient of the cosine interaction vanishes. Similarly, the composite electron tunneling amplitude is zero and so they cannot tunnel across the interface at low energies. This phase should occur when electron tunneling is a weak/irrelevant perturbation to the edge theory, and corresponds to the two sides being completely decoupled at long wavelengths. We refer to this edge phase as the \it uncoupled interface.\rm 

(2) $\overline{\Phi}_{\up\down} \neq 0$, but $\cos( 2 \phi_\up + 2\phi_\down - a_{\up\down})$ is irrelevant or marginal. 
Here, the composite electrons can now tunnel across the interface, due to the presence of the term
$\sqrt{\frac{t}{2}}\overline{\Phi}_{\up\down} ( \psi_\up^\dagger \psi_\down + {\rm h.c.})$. 
However, if $\cos( 2 \phi_\up + 2\phi_\down - a_{\up \down})$ is irrelevant or marginal, $\phi_{\uparrow}$ and $\phi_{\downarrow}$ fluctuate freely and are gapless or unconstrained at long wavelengths. Therefore, the electron correlations remain power law along the interface. This is the \it partially fused interface \rm phase. 

(3)  $\overline{\Phi}_{\up\down} \neq 0$, and both $\cos( 2 \phi_\up + 2\phi_\down - a_{\up\down})$ and $(\psi_\up^\dagger \psi_\down + {\rm h.c.})$ are relevant. Here, the composite electrons
can tunnel across the interface, and the bosons $\phi_{\up}$ and $\phi_{\down}$ are locked to one another and no longer fluctuate
freely. Consequently, the electron correlations along the interface are now qualitatively the same as in the bulk, and decay exponentially. 
This is the \it fully fused interface\rm, as the two, initially separate, CFLs have fully healed.

In principle, one might imagine a fourth phase, where  $\overline{\Phi}_{\up\down} \neq 0$, and $\cos(2 \phi_\up + 2 \phi_\down - a_{\up\down})$ is relevant, but 
the composite electron tunneling operator across the interface, $e^{i a_{\up\down}} \psi_\up^\dagger \psi_\down$, is irrelevant. 
However, since the composite electrons $\psi$ form a Fermi sea in the bulk, we expect that their tunneling across the 
interface is highly unlikely to be irrelevant. 

In addition, one may consider higher order ``partially fused'' phases, where $\overline{\Phi}_{\up\down} = 0$, but one instead decouples
pair tunneling, or higher order tunneling terms, across the interface using a Hubbard-Stratonovich transformation. These phases
would, for example, allow pairs of composite fermions to tunnel across the interface, but not single composite fermions. 

\subsubsection{Application to Pfaffian - I - Pfaffian junctions}

An interesting corollary of the above analysis appears in the case where the composite fermions of the CFL form a $p_x + i p_y$-paired superconducting state. In this case, the system then forms the famous Moore-Read Pfaffian state.\cite{Moore1991,greiter1991,read2000} A single edge of such a state consists of a chiral boson mode, $\phi$, together with a chiral Majorana mode, $\eta$. The electron tunneling term between two Moore-Read Pfaffian states 
separated by an insulating barrier takes the form:
\begin{align}
\mathcal{L}_{\text{tunn};\up\down} = t \eta_\up \eta_\down e^{2 i (\phi_\up + \phi_\down)} + {\rm h.c.}
\end{align}

This interface also supports three edge phases in principle:
\begin{enumerate}
\item \it Decoupled interface, \rm where the electron tunneling term is irrelevant. 
\item \it Partially fused interface. \rm Electron tunneling is strong/relevant. 
The Majorana tunneling operator $\eta_\up \eta_\down$ 
acquires an expectation value: $\langle \eta_\up \eta_\down \rangle \neq 0$,
which effectively gaps the counterpropagating Majorana modes. Consequently,
single electron correlations decay exponentially along the interface. 
Furthermore, the expectation value $\langle \eta_\up \eta_\down \rangle \neq 0$
allows the neutral fermion of the bulk Moore-Read state to propagate coherently across the interface. 
However, the boson tunneling term $e^{2i (\phi_\up + \phi_\down)}$ is irrelevant or marginal, so the counter-propagating chiral boson modes remain gapless along the interface. 
\item \it Fully fused interface. \rm Electron tunneling is strong/relevant and the Majorana tunneling operator $\eta_\up \eta_\down$ again acquires an
expectation value. 
But now, $e^{2i (\phi_\up + \phi_\down)}$ is relevant and pins the boson modes, causing them to acquire an energy gap. The junction is now fully gapped,
and the two sides of the interface are fully healed into each other. 
\end{enumerate}

\subsection{CFL - I - Anti-CFL Junctions}
\label{cfl-acfl}

The CFL - Insulator - Anti-CFL junction can be analyzed in a manner closely analogous to the case of the CFL - I - CFL junctions. 
In this case, we find that there are three basic types of interface phases:
\begin{enumerate}
\item \it Decoupled interface. \rm Electron tunneling between edges is weak and irrelevant. Composite electrons of the CFL cannot tunnel across the boundary into the composite holes of the Anti-CFL. Tunneling density of states decays as a power law at low energies along the interface, but exponentially in the bulk. 

\item \it Partially fused interface 1. \rm Electron tunneling across the interface is strong and/or relevant. Composite electrons of the CFL can directly tunnel across the boundary into composite holes of the Anti-CFL, but the counter-propagating bosonic edge modes remain gapless. 
Single electron correlations, and therefore the single particle tunneling density of states, have algebraic decay along the interface, but exponential decay in the bulk.
\item \it Partially fused interface 2. \rm Electron tunneling across the interface is strong and/or relevant. Composite electrons of the CFL can directly tunnel across the boundary into composite holes of the Anti-CFL, but a single chiral bosonic field remains gapless. Single electron correlations decay exponentially along the interface and the bulk. 
\end{enumerate}

The existence of the partially fused interfaces is remarkable, because the composite electrons of the CFL and the composite holes of the Anti-CFL represent distinct types of excitations of the electron system. An important difference as compared with the CFL - I - CFL junctions is the absence of a fully fused interface; the existence of at least one chiral bosonic edge field in the effective field theory suggests that the CFL and Anti-CFL are distinct phases of matter. 

In addition, we can imagine a regime of parameters where the electron tunnels across the interface via the filled Landau level of the Anti-CFL.
Such tunneling is always present, and may be irrelevant or marginal at any of the given interface phases listed above.
(If such a tunneling is relevant, presumably the system flows to one of the interfaces phases described above). 

Let us consider the case where the region $y > \epsilon/2$ is now in the Anti-CFL state. In this case, electron tunneling across the interface can occur via the terms:
\begin{align}
\label{CFLAntiCFLinterface}
\mathcal{L}_{\text{tun};\up\down} =  t_1 \chi_\uparrow \psi_\downarrow e^{2 i (\phi_{1\uparrow} + \phi_\downarrow)} + t_2 e^{ i \phi_{2\up} + 2 i\phi_\down} \psi_\down + {\rm h.c.}
\end{align}
The existence of two single-electron tunneling terms, with amplitudes $t_1$ and $t_2$, reflects the fact that the electron can be created in two distinct ways along the edge of the Anti-CFL. 

Following closely the analysis in the previous section, we can decouple the first term:
\begin{align}
\mathcal{L}_{\text{tun};\up\down} &=
\sqrt{\frac{|t_1|}{2}} \left(\Phi_{\uparrow\downarrow} \chi_\up \psi_\down + \Phi_{\uparrow\downarrow}^* e^{2i (\phi_{1\up} + \phi_\down)+i\theta} + {\rm h.c.}\right) 
\nonumber \\
&- \frac{1}{2}|\Phi_{\uparrow\downarrow}|^2 +  (t_2 e^{- i \phi_2 + 2i \phi_\down} \psi_\down + {\rm h.c.}),
\end{align}
where as before we have dropped an overall constant, absorbed one of the four-fermion terms into the density-density interactions between the separated edges, and set $t_1 = |t_1| e^{i\theta}$. Under
gauge transformations, the fields transform as:
\begin{align}
\psi_\down &\rightarrow e^{i\Lambda_\down} \psi_\down ,
\nonumber \\
\chi_\up &\rightarrow e^{i\Lambda_\up} \chi_\up,
\nonumber \\
a^\alpha &\rightarrow a^\alpha + \partial \Lambda_\alpha,
\nonumber\\
\Phi_{\up\down} &\rightarrow e^{-i \Lambda_\up - i \Lambda_\down} \Phi_{\up\down},
\nonumber \\
\phi_\down &\rightarrow \phi_\alpha - \Lambda_\down/2, \cr
\phi_{1 \up} &\rightarrow \phi_{1\up} - \Lambda_\up/2,
\end{align}
where $\alpha = \up$ or $\down$. 

As before, we consider a self-consistent mean-field solution 
\begin{align}
\langle \Phi_{\up \down} \rangle = \sqrt{2|t|} \langle \chi_\up \psi_\down + e^{-2 i (\phi_{1\up} + \phi_\down + \theta)} \rangle.
\end{align}
To understand the physics when $\overline{\Phi}_{\up \down}$ is non-zero, we expand the action around this mean-field solution by setting:
\begin{align}
\Phi_{\up\down} = \bar{\Phi}_{\up\down} e^{i a_{\up\down}}. 
\end{align}
This leads to the following interface Lagrangian at long wavelengths:
\begin{align}
\label{cflacfltunn}
\mathcal{L}_{\text{tun};\up\down} = &\kappa \overline{\Phi}_{\up\down}^2 (a^{\up} + a^{\down} + \partial a_{\up\down})^2 +
\sqrt{\frac{|t|}{2}}\overline{\Phi}_{\up\down} ( e^{i a_{\up\down}} \chi_\up \psi_\down
\nonumber \\
& + {\rm h.c.} + \cos( 2 \phi_{1\up} + 2\phi_\down - a_{\up\down})),
\end{align}
where $a_{\up \down} \rightarrow a_{\up \down} - \Lambda_\up - \Lambda_\down$ under local gauge transformations.
As before, we see that when $\bar{\Phi}_{\up\down} = 0$, the two states are effectively decoupled. 
When $\bar{\Phi}_{\up\down} \neq 0$, we see that it is possible for the composite electron $\psi$ to tunnel across
the interface directly into the composite hole $\chi$. 
Importantly, since the edge boson fields have a net chirality, they cannot be fully gapped by any
type of backscattering term, and therefore the existence of at least one gapless edge field is guaranteed. 

In order to understand the possible phases for the edge boson fields, let us consider their commutation relations:
\begin{align}
[\phi_{\down} (x), \phi_{\down}(y)] = i \frac{\pi}{2} \text{sgn}(x-y),
\nonumber \\
[\phi_{1\up} (x), \phi_{1\up}(y)] = i \frac{\pi}{2} \text{sgn}(x-y),
\nonumber \\
[\phi_{2\up} (x), \phi_{2\up} (y)] = -i \pi \text{sgn}(x-y) . 
\end{align}
These commutation relations imply that the cosine term $\cos( 2 \phi_{1\up} + 2\phi_\down - a_{\up\down}) )$ in Eqn. (\ref{cflacfltunn})
cannot pin its argument to a constant value in space, as the argument does not commute with itself at different points in space. 
Therefore, we see the appearance of an interface phase that we refer to as \it partially fused interface 1\rm. The composite holes
can directly tunnel across the interface into composite electrons, because $\bar{\Phi}_{\up\down} \neq 0$, and the single electron
correlations $\langle c^\dagger(x,t) c(0,0) \rangle$ decay algebraically because all of the interface fields are gapless. 

The existence of three gapless interface fields, $\phi_\down, \phi_{1\up}, \phi_{2\up}$ is not guaranteed. 
Within a partially fused phase, a term of the form,
\begin{align}
\label{gappingTerm}
\cos(-2 \phi_{1\up} + 2\phi_\down - 2\phi_{2\up} - a_{\up\down}'),
\end{align}
can pin its argument to a constant value in space, thus effectively eliminating the right-moving chiral field
$\phi_{1 \up} - \phi_\down$ and left-moving chiral field $\phi_{2 \up}$ from the low energy physics, leaving behind a
single right-moving chiral field, 
\begin{align}
\phi_n = \phi_{1\up} + \phi_\down.
\end{align}
The cosine term, Eqn. (\ref{gappingTerm}), can be generated in the effective action as follows. We consider a correlated 
electron tunneling term across the interface:
\begin{align}
\mathcal{L}_{\text{tunn};\up\down}^{'} &= t_3 c_\down^\dagger c_{1\up}^\dagger c_{2\up} \partial_x c_{2\up} + {\rm h.c.}
\nonumber \\
&= t_3 \psi_\down^\dagger \chi_\up e^{-i 2\phi_\down + i 2 \phi_{1\up} + 2 i\phi_{2\up}} + {\rm h.c.}
\end{align}
Decoupling the fermion and boson fields with a Hubbard-Stratonovich transformation, as before, we obtain:
\begin{align}
\mathcal{L}_{\text{tunn};\up\down}' &= 
\sqrt{\frac{t_3}{2}} ( \psi_\down^\dagger \chi_\up \Theta_{\up\down} + 
\nonumber \\
&\Theta_{\up\down}^* e^{-i 2\phi_\down + i 2 \phi_{1\up} + 2 i\phi_{2\up}} + {\rm h.c.} )
- |\Theta_{\up\down}|^2/2
\end{align}
Expanding $\Theta_{\up\down}$ around its mean-field value $\overline{\Theta}_{\up\down}$ by including its phase fluctuations:
\begin{align}
\Theta_{\up\down} = \overline{\Theta}_{\up\down} e^{i a'_{\up\down}} , 
\end{align}
we obtain 
\begin{align}
\mathcal{L}_{\text{tunn};\up\down}' = &
\sqrt{\frac{t_3}{2}} ( 
\psi_\down^\dagger \chi_\up \overline{\Theta}_{\up\down} e^{i a_{\up\down}'} + 
\nonumber \\
&\overline{\Theta}_{\up\down} \cos(- 2\phi_\down + 2 \phi_{1\up} + 2 \phi_{2\up} - a_{\up\down}') 
+ {\rm h.c.} 
)
\end{align}
Therefore, when $\overline{\Theta}_{\up\down} \neq 0$ and the cosine term above is relevant, the the field $-\phi_\down + \phi_{1\up} + \phi_{2\up} - a'_{\up \down}$ is pinned. 

In order to understand the physical consequence of this, observe that none of the electron operators, $c_{1\up}, c_{2\up}, c_\down$ commute with the 
argument of the cosine, $- 2\phi_\down + 2 \phi_{1\up} + 2 \phi_2 - a_{\up\down}'$. In fact, each of the electron creation operators, $c_\alpha^\dagger(x)$, 
for $\alpha,\beta = \down, 1\up,$ and $2\up$, create a $2\pi$ kink in the field $- 2\phi_\down + 2 \phi_{1\up} + 2 \phi_{2\up} - a_{\up\down}'$. 
The operator $c_\alpha^\dagger(x) c^\dagger_\beta(x')$, for $\alpha,\beta = \down, 1\up,$ and $2\up$, when acting on the ground state of the system,
creates a finite energy pair of domain walls in $- 2\phi_\down + 2 \phi_{1\up} + 2 \phi_{2\up} - a_{\up\down}'$. Therefore, 
single electron correlations must decay exponentially along the interface. In fact, correlation functions of all operators that carry non-zero electric charge
must, for the same reason, decay exponentially along the interface. However, since the field $\phi_{1\up}+\phi_\down$ is gapless, electrically neutral
operators, such as $c_\down^\dagger c_{1\up}$, can have power-law correlations along the boundary, just as other neutral operators do in the bulk. 

The interface theory can then be written as:
\begin{widetext}
\begin{align}
\mathcal{L}_{\text{edge}} &= \frac{1}{4\pi}[ (\partial_t \phi_n - a_t^\up - a_t^\down)\partial_x \phi_n + \frac{1}{4} a_t^\alpha a_x^\alpha
- v_n  (\partial_x \phi_n - \frac{1}{2} a_x^\up - \frac{1}{2} a_x^\down)^2]
\nonumber \\
&+ \kappa \overline{\Phi}_{\up\down}^2 (a^{\up} + a^{\down} + \partial a_{\up\down})^2 +
\sqrt{\frac{|t|}{2}}\overline{\Phi}_{\up\down} ( \chi_\up \psi_\down e^{i a_{\up\down}} + {\rm h.c.} )
\nonumber \\
&+ \kappa' \overline{\Theta}_{\up\down}^2  (a^{\up} - a^{\down} + \partial a_{\up\down}')^2 +
\sqrt{\frac{t_3}{2}} \overline{\Theta}_{\up\down} ( \psi_\down^\dagger \chi_\up e^{i a_{\up\down}'} + {\rm h.c.}).
\end{align}
\end{widetext}

It is useful to note that the $\nu =1$ neutral mode $\phi_n$ at this interface can be fermionized by introducing a chiral fermion field:
\begin{align}
\Psi_R \sim e^{i (\phi_{1\up} + \phi_\down)} .  
\end{align}
We will revisit this point in Sec. \ref{trans} to understand the emergence of a Dirac fermion at the quantum phase transition between the CFL and Anti-CFL. 

Therefore, as summarized at the beginning of this section, the CFL - I - Anti-CFL interface hosts three distinct interface phases, two of which allow topological transmutation of composite electrons of the CFL into composite holes of the Anti-CFL. 

\subsubsection{Application to Pfaffian - I -Anti-Pfaffian junctions}

As in the CFL - I - CFL case, an interesting corollary of the above analysis appears in the case where the composite fermions of the CFL condense into a $p_x + i p_y$ paired state, while the composite holes of the Anti-CFL condense into a $p_x - i p_y$ paired state. In this case, the CFL is replaced by the Moore-Read Pfaffian state, while the Anti-CFL is replaced by the Anti-Pfaffian. 

The boundary between the Pfaffian and Anti-Pfaffian is described by the Lagrangian
\begin{align}
\mathcal{L} = &- \frac{1}{4\pi} K_{IJ} \partial_t \phi_I \partial_x \phi_J - V_{IJ} \partial_x \phi_I \partial_x \phi_J ,
\nonumber \\
&+ i \eta_\alpha (\partial_t - v \partial_x) \eta_\alpha
\end{align}
with $K = \left(\begin{matrix} 2 & 0 & 0 \\ 0 & -1 & 0 \\ 0 & 0 & 2 \end{matrix} \right)$, and where $\eta_\alpha$ are chiral Majorana fields, each propagating with velocity $v$. 
For convenience, we have relabelled the scalar fields $\phi_1 = \phi_{1\up}$, $\phi_2 = \phi_{2\up}$, $\phi_3 = \phi_\down$.

The considerations above imply the existence of several distinct interface phases at the junction between the Pfaffian and Anti-Pfaffian:
\begin{enumerate}
\item \it Decoupled interface\rm, where electron tunneling across the interface is irrelevant. 
\item \it Partially fused interface 1\rm. The copropagating Majorana modes spontaneously acquire an expectation value,
$\langle \eta_\up \eta_\down \rangle \neq 0$, which does not cause them to acquire an energy gap, but which does
allow the neutral fermion of the bulk Pfaffian state to directly tunnel across the interface as the neutral fermion of the bulk Anti-Pfaffian state.
Single electron correlations remain algebraic along the interface. 
The interface has a chiral central charge $c = c_L - c_R = -1 + 3 = 2$, where $c_L = 1$ from the left-moving $\nu=1$ mode, and $c_R = 3$ from the 
two right-moving Majorana fermion modes and two right-moving $\nu=1/2$ boson modes. 
\item \it Partially fused interface 2\rm. The three bosonic modes are gapped, due to a large tunneling term,
$\cos(-4 \phi_1 + 4\phi_2 + 4\phi_3)$, leaving behind a single right-moving neutral $\nu = 1$ boson mode,
$\phi_n = \phi_1 + \phi_2$.\footnote{A similar set of boson modes and gapping term was also considered recently in Ref. \onlinecite{mross2014}, in a different context.}  
Correlation functions of charged operators, such as single-electron correlations, therefore decay exponentially along the interface. The neutral fermion of the Pfaffian cannot tunnel
across into the neutral fermion of the Anti-Pfaffian because $\langle \eta_\up \eta_\down \rangle = 0$. 
\item \it Partially fused interface 3\rm. Same as partially fused interface 2, except that $\langle \eta_\up \eta_\down \rangle \neq 0$, so that the neutral fermion from the Pfaffian can tunnel into the neutral fermion of the Anti-Pfaffian. 
\end{enumerate}
In all of these cases, the interface hosts edge modes that are stable to local perturbations. 
This provides a non-trivial example where the boundary between two topological phases possesses robust gapless chiral edge modes, but nevertheless
hosts several different topologically distinct boundary phases. Similar examples of topologically distinct boundary phases for gapped interfaces of Abelian topological states 
have been classified recently in Ref. \onlinecite{levin2013,barkeshli2013defect,barkeshli2013defect2}, and interesting examples of topologically distinct gapless boundary phases
have recently been discovered in Ref. \onlinecite{cano2014}.

\section{Continuous Transition between CFL and Anti-CFL}
\label{trans}

In the previous sections, we have developed a bulk effective field theory for the Anti-CFL. 
A natural question now is to understand the nature of the zero temperature phase transition 
between the CFL and Anti-CFL. This can be tuned, for example, by varying the filling fraction through $\nu = 1/2$ in the limit of zero Landau level mixing. 
One can distinguish two possibilities, depending on whether the phase transition is first order or continuous in the clean (disorder-free) limit. 
If first order, as we will briefly describe, the transition will be rendered continuous in the presence of disorder. 
A continuous zero temperature quantum phase transition will have important consequences for broad regions of the finite temperature phase diagram
in the vicinity of $\nu = 1/2$.

\subsection{Clean Critical Point}

In this section, we present a theory that describes a continuous transition between the CFL and Anti-CFL in the absence of disorder.

Recall the effective action in Eqn. (\ref{antiCFL2}) for the Anti-CFL. Let us rewrite this action by introducing the linear combinations:
\begin{align}
a^\rho &= -\tilde{a} - A,
\nonumber \\
a^\sigma &= 2\tilde{a} + A.
\end{align}
In terms of these gauge fields, the effective action for the Anti-CFL becomes
\begin{align}
\mathcal{L}_{\rm ACFL} =& -\frac{2}{4\pi} a^\rho \partial a^\rho + \frac{1}{2\pi} (a - A_E) \partial a^\rho 
\nonumber \\
&+ \frac{1}{2\pi} a \partial a^\sigma +\frac{1}{4\pi} a^\sigma \partial a^\sigma +
\nonumber \\
&\chi^\dagger (i \partial_t + a_t + \mu_h) \chi - \frac{1}{2m_h} \chi^\dagger (i \partial_i + a_i)^2 \chi .
\end{align}
Now we can see that if $a^\sigma$ were set to zero, this would be identical to $\mathcal{L}_{\text{CFL}}$ (see Eqn. (\ref{partonCFL2}), 
with the replacement of $\chi$ by $\psi$, $m_h$ by $m_e$, and $\mu_h$ by $\mu_e$). 
This motivates the following theory, which can describe the transition between CFL and Anti-CFL:
\begin{align}
\label{transitionB}
\mathcal{L}_{\text{trans}} &= -\frac{2}{4\pi} a^\rho \partial a^\rho + \frac{1}{2\pi} (a - A_E) \partial a^\rho 
\nonumber \\
&+ \frac{1}{2\pi} a \partial a^\sigma +\frac{1}{4\pi} a^\sigma \partial a^\sigma
+ \chi^\dagger (i \partial_t + a_t + \mu) \chi 
\nonumber \\
& - \frac{1}{2m} \chi^\dagger (i \partial_i + a_i)^2 \chi + |(i \partial + a^\sigma) \varphi|^2 + V(\varphi) . 
\end{align}
When $\varphi$ is disordered, $\langle |\varphi| \rangle = 0$, this theory describes the Anti-CFL state. 
When $\varphi$ orders, $\langle |\varphi| \rangle \neq 0$, $a^\sigma$ gets Higgsed. Setting $a^\sigma = 0$ recovers $\mathcal{L}_{{\rm CFL}}$. 
At the transition, $\mu_e = \mu_h = \mu$ and $m_\psi = m_\chi = m$; these parameters may vary away from the transition from their 
values at the transition.  

\begin{figure}
	\centering
	\includegraphics[width=2.0in]{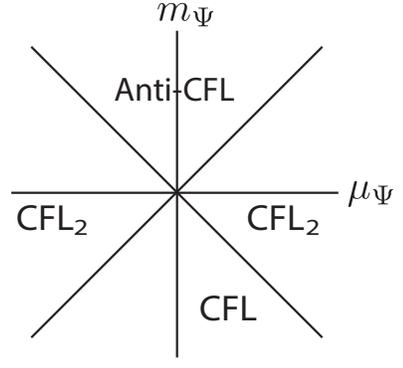}
	\caption{\label{zeroTPD} Zero temperature (mean-field) phase diagram of Eqn. (\ref{transitionF}).}
\end{figure}

Examining Eqn. (\ref{transitionB}), we see that the CS term for $a^\sigma$ attaches a unit flux to $\varphi$ to convert it to a fermion. In fact, using the duality proposed in Ref. \onlinecite{chen1993}, the boson $\varphi$ can be fermionized, yielding the following dual effective theory of the critical point:
\begin{align}
\label{transitionF}
\mathcal{L}_{\text{trans}} &= -\frac{2}{4\pi} a^\rho \partial a^\rho + \frac{1}{2\pi} (a - A^E) \partial a^\rho - \frac{1}{2} \frac{1}{4\pi} a \partial a
\nonumber \\
&+ \chi^\dagger (i \partial_t + a_t + \mu) \chi - \frac{1}{2m} \chi^\dagger (i \partial_i + a_i)^2 \chi 
\nonumber \\
&+\Psi^\dagger \sigma_t [\sigma_\mu (i \partial_\mu + a_\mu) - m_\Psi ] \Psi + \mu_\Psi \Psi^\dagger \Psi ,
\end{align}
where $\Psi$ is a two-component Dirac fermion, and $\sigma_t \equiv \sigma_z$, $\sigma_x, \sigma_y$ are the Pauli matrices. We have included
a chemical potential term $\mu_\Psi$, which is allowed in general; the analogous term in Eqn. (\ref{transitionB}) is difficult to analyze because it involves
monopole operators. When $m_\Psi < 0$ and $|m_\Psi| > |\mu_\Psi|$, integrating out 
$\Psi$ cancels the CS term for $a$, and gives the CFL theory, $\mathcal{L}_{\text{CFL}}$. When $m_\Psi > 0$ and $|m_{\Psi}| > |\mu_\Psi|$, 
integrating out $\Psi$ gives an additional $-\frac{1}{2} \frac{1}{4\pi} a \partial a$. The resulting Lagrangian can be seen to be equivalent to that of 
the Anti-CFL theory. 

When $|\mu_\Psi| > |m_\Psi|$, the $\Psi$ fermions form a Fermi sea, and the system transitions into a novel type of CFL-like state but with
two bands of composite-fermion-type excitations, which we label CFL$_2$. The phase diagram described above is sketched in Fig. \ref{zeroTPD}. 

Importantly, we see that within this theory, the only way to tune from the CFL to the Anti-CFL states is to pass through a phase transition. 
There is a direct phase transition when $\mu_\Psi = 0$; when $\mu_\Psi \neq 0$, there is an intermediate phase. In the following discussion, we analyze the direct 
transition as $m_\Psi$ is tuned through zero, with the chemical potential fixed at $\mu_\Psi = 0$. 

\subsubsection{Coupled Chain Description}

\begin{figure}
	\centering
	\includegraphics[width=2.0in]{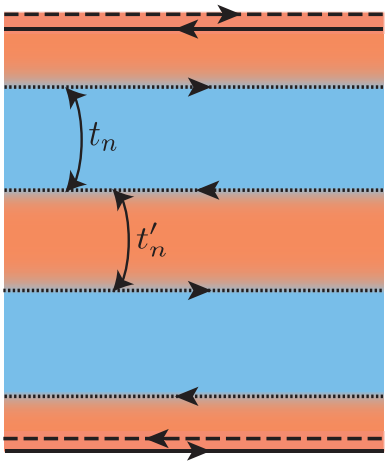}
	\caption{\label{coupledChain} The transition between the CFL and Anti-CFL can be understood by considering alternating strips of the two phases. 
$t_n$ and $t_n'$ describe the backscattering of counterpropagating chiral fermions, as shown. }
\end{figure}
An interesting way to approach the critical point, which makes contact to the results of Sec. \ref{cfl-acfl}, is to consider alternating 
strips of CFL and Anti-CFL, reminiscent of the percolation picture for the IQH plateau transition of Chalker and Coddington.\cite{chalkercoddington}
In Sec. \ref{cfl-acfl}, we showed that the interface between the CFL and Anti-CFL necessarily possesses
a chiral electrically neutral fermion field (or, alternatively, a $\nu=1$ neutral chiral boson field). Therefore, alternating strips of CFL and
Anti-CFL possess counterpropagating chiral fermion fields, as shown in Fig. \ref{coupledChain}. For a finite size strip, there will be 
backscattering terms between these counterpropagating fields, which we have labelled $t_n$ and $t_n'$ in Fig. \ref{coupledChain}. 
When $t_n > t_n'$, the system is in the Anti-CFL state everywhere. When $t_n' > t_n$, the system is in the CFL state everywhere;
it can be easily verified that the remaining edge fields can be reduced, via backscattering terms, to the minimal CFL boundary
theory introduced in Sec. \ref{edgeTheories}. At the critical point, $t_n = t_n'$, the interface fields delocalize in both directions and
form the 2+1D Dirac fermion field $\Psi$. 

\subsubsection{Mean-field Treatment}

The transition at $m_\Psi = 0$ is continuous in the absence of gauge fluctuations of $a$, if the Dirac fermion $\Psi$ 
does not couple to the holes $\chi$. 
In what follows, we show that the transition remains continuous in the presence of these 
fluctuations as well. 
The analysis below extends previous work in Ref. \onlinecite{senthil2008} for the 
continuous Mott transition between a Landau Fermi liquid and a $U(1)$ spin liquid Mott insulator, and also the closely related analysis of Ref. \onlinecite{barkeshli2012hlr}
for the transition between the CFL and a Landau Fermi liquid. 

First, let us consider the continuity of the transition in a mean-field limit where the gauge fluctuations of $a$ are ignored. 
In this limit, we must analyze the effects of any coupling between the Dirac fermion $\Psi$ and hole $\chi$.
In principle, there are three additional quadratic interactions that one can add to Eqn. (\ref{transitionF}),
\begin{align}
{\cal L}_{\rm quad} = g_1 \chi^\dagger \chi + (g_2)_\alpha (\Psi^\dagger_\alpha \chi + {\rm h.c.}), 
\end{align}
for $\alpha = 1,2$, that may be added to ${\cal L}_{\rm trans}$ in Eqn. (\ref{transitionF}).
The first interaction parameterized by $g_1$ is merely a shift of the hole chemical potential which is fixed so that the system sits at half-filling.
The second interaction parameterized by $g_2$ mixes the $\Psi$ and $\chi$ fields.
In the clean limit, this interaction vanishes at long wavelengths due to an oscillating coupling, proportional to $\exp(i k_F x)$, arising from the hole Fermi sea.
As described in the previous subsection, we tune the $\Psi$ chemical potential $\mu_\Psi \rightarrow 0$.

We now consider quartic interactions between the Dirac fermion
current and the particle-hole continuum of the $\chi$ fermions:
\begin{align}
\int dt d^2 x\Big[\lambda_\mu j^\Psi_\mu \chi^\dagger \chi\Big],
\end{align} 
where $j^\Psi_\mu =  \Psi^\dagger \sigma_\mu \Psi$ and $\lambda_\mu$ are coupling constants.

Strict perturbation theory about the decoupled fixed point indicates that the interactions parameterized by $\lambda_\mu$ are irrelevant.
The tree-level momentum-space scaling assignments are as follows: $[\omega] = 1, [k_\perp]=1, [k_\parallel] = 0, [p_\mu] = 1, [\chi] = -3/2$, and $[\Psi]=-2$, where $k_\perp$ refers to the momentum of the composite hole perpendicular to the Fermi surface, $k_\parallel$ is the momentum parallel to the Fermi surface, and $p_\mu$ collectively denotes the frequency and momentum of the Dirac fermion. For generic values of the hole momentum, the interaction has momentum-space scaling dimension equal to $+2$, while for special kinematic values, say, when the incoming and outgoing hole have equal and opposite momentum, the interaction is dimension $+1$. 
In both cases, the interaction is irrelevant at weak coupling.

We may also consider the interaction generated after integrating out the hole Fermi sea.
Integrating out the excitations about the hole Fermi surface leads to an interaction of the form
\begin{align}
\label{RPApert}
\int d\omega d^2 k \Pi_b(\omega, k) |\lambda_\mu j_{\mu}^\Psi (k, \omega) |^2,
\end{align}
where $\Pi_{b}(\omega, k) \approx c_1 + \frac{c_2 |\omega|}{k}$ in the limit that $\omega/k \rightarrow 0$. 
Within this mean-field treatment,
$c_1$ is the finite, non-zero compressibility of the $\chi$ fermions while $c_2$ is a second finite constant. 
The operator $j^\Psi_{\mu}(\omega, k)$ has (momentum-space) scaling dimension $-1$ about the Dirac fermion critical point where 
the dynamic critical exponent $z=1$, implying that (\ref{RPApert}) is an irrelevant perturbation. 
Therefore, the Dirac fermion $\Psi$ and composite hole $\chi$ are decoupled at long wavelengths in the absence of the fluctuations of the 
gauge field $a$, thereby, implying that in mean-field theory, the transition remains continuous.

\subsubsection{Gauge Fluctuations}

Next, let us discuss a few properties of the transition in the presence of fluctuations of the gauge field $a$. 
We show that the transition remains continuous in the presence of the gauge fluctuations, which leads to a rich finite 
temperature phase diagram with two finite temperature crossover scales.

We shall be interested in the behavior of the theory in the kinematic regime, $0 \leq |\omega| < v_F |k| \ll v_F k_F$, where $\omega$ refers to the Wick-rotated (imaginary time) frequency, $k_F = \sqrt{2 m \mu}$ is the Fermi momentum at half-filling, and the Fermi velocity $v_F = k_F/m$.
We treat the problem using the random phase approximation (RPA).

We work in Coulomb gauge, $\partial_i a_i = 0$. 
The temporal gauge fluctuations of $a_0$ are screened by the $\chi$ fermions, and decouple from the low-energy physics. 
The transverse magnetic fluctuations of $a_i$ are damped by both $\Psi$ and $\chi$ fluctuations.
In the kinematic regime, $|\omega| \sim |k| \ll k_F$, $a_i$ first receives the one-loop self-energy correction due to fluctuations of the $\Psi$ fermions,
\begin{align}
\label{diraccontribution}
\Pi^{(\Psi)}_{i j}(\omega, k_i) = {\sqrt{\omega^2 + |k|^2} \over 16} \delta_{i j}.
\end{align}
(Here, we have identified the cutoff of the entire theory with the Fermi wave vector.)
The correction in Eqn. (\ref{diraccontribution}) softens the IR behavior of the $a_i$ propagator.
In the RPA, the $a_i$ propagator is diagonal: 
\begin{align}
\label{qed3prop}
G_{ij}(\omega \sim |k|) = {16 \over \sqrt{\omega^2 + |k|^2}} \delta_{ij}.
\end{align}
While not present for $m_\Psi = 0$, an off-diagonal term arising from a bare Chern-Simons coupling is subleading in the RPA.
We note that the transverse gauge field propagator in Eqn. (\ref{qed3prop}) is identical to the one-loop corrected gauge field propagator of QED3.

As we proceed to lower frequencies $0 \leq |\omega| < v_F |k|$, the composite hole sea Landau damps the transverse gauge field leading to an additional (diagonal) self-energy correction,
\begin{align}
\label{holecontribution}
\Pi_{i j}^{(\chi)} = \gamma_0 {|\omega| \over |k|} \delta_{ij},
\end{align}
where $\gamma_0 = {m v_F \over 2\pi}$.
Putting together these two self-energy corrections in Eqns. (\ref{diraccontribution}) and (\ref{holecontribution}), the gauge field propagator takes the asymptotic form:
\begin{align}
\label{gaugeasymp}
G_{ij}(\omega < v_F |k| \ll v_F k_F) = {\delta_{i j} \over \gamma_0 {|\omega| \over |k|} + {|k| \over 16}}.
\end{align}

To understand whether or not the Dirac fermions remain massless at the transition in the presence of gauge fluctuations, we need to determine the form of the Dirac fermion self-energy $\Sigma_\Psi$ at the transition.
If this self-energy approaches a finite, non-zero constant at low frequencies and momenta, then we conclude that the Dirac fermions obtain a mass at the transition, thereby, implying a fluctuation-driven first-order transition; otherwise, the transition remains continuous in the presence of interactions.
Using the asymptotic form for the gauge field propagator in Eqn. (\ref{gaugeasymp}), the one-loop correction to $\Sigma_\Psi$ in the regime $|\omega| < v_F |k|$ takes the form:
\begin{widetext}
\begin{align}
\Sigma_{\Psi}(q_0, |q|) = \int {d \omega d^2 k \over (2 \pi)^3} \Big({(\omega + q_0)\sigma^z + (k_i + q_i) \sigma^i \over (\omega + q_0)^2 + |k + q|^2} \Big) \Big({|k| \over \gamma_0 |\omega| + {k^2 \over 16}} \Big).
\end{align}
\end{widetext}
At $q_0 = |q| = 0$, we find that the above self-energy vanishes.
This indicates that the transition remains continuous at leading order.
Intuitively, we note that the $|\omega|/q$ contribution to the propagator for the transverse gauge fluctuations
looks, from the point of view of the Dirac fermions, like a mass term for the gauge field, because of the relativistic ($z = 1$)
dispersion of the Dirac fermions. This provides another way to see that the damped gauge fluctuations 
should continue to leave the Dirac fermion transition continuous. 
 
We comment that the composite holes receive a self-energy correction of the marginal Fermi liquid\cite{MFL} 
form due to their interaction with the transverse gauge field precisely at the transition point.
This arises from a combination of screening of the gauge field by the Dirac fermion and Landau damping by the composite hole Fermi surface, i.e., the gauge field propagator in Eqn. (\ref{gaugeasymp}). In contrast, the scaling of the self-energy of the composite holes or composite electrons on the $m_\Psi>0$ or $m_\Psi<0$ side of the transition depends upon the form of the electron-electron density-density interaction.
As we remark below, if the Coulomb interaction is un-screened, the self-energy has a marginal Fermi liquid form on both sides of the transition as well. For shorter-ranged interactions, the frequency-dependent self-energy is expected to scale with a power that is less than unity.

\subsubsection{Finite-Temperature Phase Diagram}

It is interesting to examine the effect of the fermions on the gauge field more closely.\cite{senthil2008}
On the Anti-CFL side of the transition, $m_\Psi>0$, the action for the transverse component of the gauge field, after integrating out both the Dirac fermion and composite hole, contains the terms:
\begin{align}
\label{gaugeaction}
S_{\text{eff}}[a] = & \int d\omega d^2k \Big[ \gamma_0 \frac{|\omega|}{k} + \chi_0 k^2 \cr
+ & \sigma_0 \sqrt{\omega^2 + k^2} F\Big({\sqrt{\omega^2 + k^2} \over |m_\Psi|}\Big) |a(\omega,k)|^2\Big],
\end{align}
where we have assumed $0 \leq \omega < v_F |k|$.
The scaling function:
\begin{align}
F(x) = & x, {\rm for}\ x =0, \cr
= & 1, {\rm for}\ x=\infty.
\end{align}
The parameters $\chi_0$, and $\sigma_0$ are finite constants.
We implement the $0 \leq |\omega| < v_F |k|$ limit by replacing:
\begin{align}
\label{approx}
\sqrt{\omega^2 + k^2} F\Big({\sqrt{\omega^2 + k^2} \over |m_\Psi|}\Big) \rightarrow  |k| F(|k|/|m_\Psi|).
\end{align}
An interesting feature of this transition is the existence of two crossover temperature scales which follows from Eqns. (\ref{gaugeaction}) and (\ref{approx}). 
For $|q| > T \sim \omega > T^* \propto m_\Psi$, the Dirac fermions are effectively massless, and therefore the system is effectively a ``quantum critical'' non-Fermi liquid. 
As the temperature is lowered below $T^\ast$, the Dirac fermions begin to decouple, however, their influence on the gauge fields continues to be felt.
The frequency and momenta of the gauge field are related by $\omega \sim (\sigma_0/\gamma_0) |k|^2 < |k|$ in the high temperature limit and so the scaling function only crosses over to its low-energy limit for $T \sim T^{**} = (\sigma_0/\gamma_0) m_\Psi^2$, which in the region about the critical point is lower than $T^\ast$.
In the crossover regime $T^{**} < T < T^*$, the Dirac fermions are effectively massive, but the gauge propagator has not changed significantly. This means that the theory is formally similar to the case where the holes $\chi$ feel a long-ranged Coulomb interaction, in which case they form a ``marginal'' composite Fermi liquid. In this marginal Anti-CFL, the self-energy of the holes scales as:
\begin{align}
\Sigma(\omega) \sim \omega \ln \omega,
\end{align}
while the specific heat scales like,
\begin{align}
C_v \sim T \ln T. 
\end{align}

For $T < T^{**}$, the fate of the system depends on whether there are long-ranged Coulomb interactions, or if they have been screened (by, e.g., an external gate). In the former case, there is no qualitative modification of these scaling forms upon lowering the temperature through $T^{**}$. In the latter case, for $T < T^*$ the fermions have a different self-energy:
\begin{align}
\Sigma(\omega) \sim \omega^{2/3},
\end{align}
for short-ranged interactions, while the specific heat scales like
\begin{align}
C_v \sim T^{2/3}. 
\end{align}
Identical considerations apply on the CFL side ($m_\Psi<0$) of the transition.

In addition, the system appears to be incompressible at the critical point.
We find the (DC) compressibility to vanish linearly as the wave vector 
$k \rightarrow 0$.

\begin{figure}
	\centering
	\includegraphics[width=3.2in]{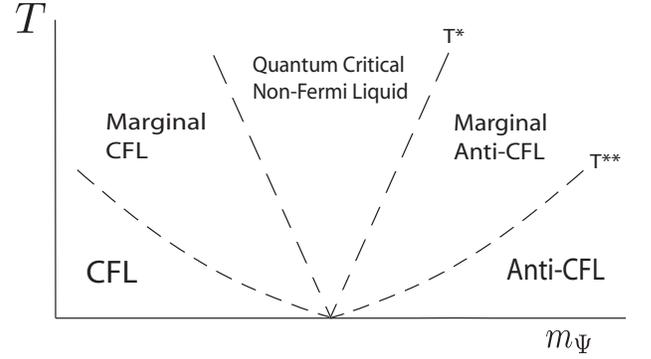}
	\caption{
\label{cflacflpd} Finite temperature phase diagram for the CFL - Anti-CFl transition, as predicted by the theory
presented in Eq. (\ref{transitionB}) - (\ref{transitionF}). 
For short-ranged interactions, the system exhibits two crossover temperature 
scales, $T^*$ and $T^{**}$. For long-ranged interactions, $T^{**}$ has no effect, and there is effectively a single crossover temperature scale $T^*$.  
The tuning parameter $m_\Psi$ shown in the figure is the Dirac fermion mass in Eqn (\ref{transitionF}), which is a phenomenological parameter in the
field theory. Physically, it can be taken to be a perturbation that breaks particle-hole symmetry and thus favors the CFL or Anti-CFL. 
}
\end{figure}

\subsection{Disordered Critical Point}

In the presence of quenched disorder (which breaks particle-hole symmetry), the transition between the CFL and Anti-CFL must be continuous. 
This follows from the absence of first-order transitions in the presence of disorder in two spatial dimensions or below.\cite{imry1979,hui1989,aizenman1989}

There are two possible cases to consider.
First, suppose that in the clean limit, the transition is continuous, possibly described by the effective field theory presented above. The effect of weak disorder can then be 
included as a perturbation, which maps onto a random mass in Eqn. (\ref{transitionF}). Since Eqn. (\ref{transitionF}) represents a non-trivial strongly coupled field theory, it is difficult
to determine the exact scaling dimensions of operators, and thus to deduce the effect of disorder. If disorder is weak and perturbatively irrelevant, then the clean critical theory will continue to 
describe the long wavelength behavior of the system down to zero temperature. If disorder is relevant, then the clean critical point would only be valid down to some crossover temperature scale,
below which the long wavelength properties will be described by an alternative disordered critical fixed point, the study of which we leave for future work. 

Suppose instead that the transition is first order in the clean limit. There is a standard argument for the rounding of a first order transition in the presence of disorder that locally favors one phase or another.\cite{imry1979,hui1989,aizenman1989} Here we will briefly review this argument, in the context of the CFL - Anti-CFL transition. If the CFL - Anti-CFL transition is first order in the absence of disorder, there is a characteristic correlation length, $\xi$, near the transition. $\xi$ can be thought of as the length scale beyond which it makes sense to identify the system as being in the CFL or Anti-CFL. It should be equivalent to the characteristic length scale of the exponential decay of bulk single electron correlations in the CFL or Anti-CFL. Thus, in the presence of disorder, the system can be broken up into regions of linear dimension $\propto \xi$, in which case it will locally be in the CFL or Anti-CFL state. Whether the system is locally in the CFL or Anti-CFL state in any given region can be considered to be an effective Ising degree of freedom. The surface tension between different neighboring regions is proportional to the length of their boundary, and can be thought of as an antiferromagnetic interaction between the effective Ising degrees of freedom. The disorder thus acts effectively like a random field in an Ising model.
However, this model does not have a phase transition in two spatial dimensions. Since we know there must be a phase transition between the CFL and Anti-CFL, this implies that the starting assumption that the transition is first order must be incorrect, and the system must instead exhibit a continuous quantum phase transition. 

We note that since the electron density does not jump across the phase transition between the CFL and Anti-CFL, long-range interactions alone, in the absence of disorder, would not be expected to round an otherwise first-order transition. 

\section{Possible experimental signatures}
\label{exp}

In this section, we consider a number of observable signatures that can distinguish the CFL and Anti-CFL. 
We reinstate the constants $\hbar = h/2\pi$, the electron charge $-e <0$, the speed of light $c$, and 
the unit of flux $\phi_0 = hc/e$. 

\subsection{DC transport: resistivity jumps}
\label{resistivitySec}

Here, we show that within the standard RPA treatment of composite fermion theory, the difference between the CFL and Anti-CFL is manifest in 
subtle differences in the low-temperature DC transport properties of the two states exactly at and near $\nu = 1/2$. 
If we assume that the system is uniformly in the CFL state for $\nu < 1/2$ and uniformly in the
Anti-CFL state for $\nu > 1/2$, and that there is a first order transition between the two states at exactly $\nu = 1/2$, then we predict jump discontinuities in the Hall and longitudinal resistivities. The DC transport can therefore provide an indirect confirmation of the existence of the 
Anti-CFL state and also provide evidence of a first order transition between the the CFL and Anti-CFL states. Absence of these discontinuities, in the limit of zero Landau level mixing, would then presumably imply a continuous transition between the CFL and Anti-CFL.

\subsubsection{CFL}

In the CFL state, the electrical resistivity is given by the Ioffe-Larkin composition rule:\cite{IoffeLarkin89}
\begin{align}
\label{CFLcomposition}
\rho^{\rm CFL} = \rho^{\text{cf}} + \rho^{\text{cs}},
\end{align}
where $\rho^{\text{cf}}$ is the contribution to the resistivity coming from the composite electrons (usually referred to as the composite fermions), and $\rho^{cs}$ is the contribution from the Chern-Simons term of the statistical gauge field representing the attached flux tubes, with
\begin{align}
\label{rhoCS}
\rho^{\text{cs}} = \frac{h}{e^2}\left( \begin{matrix} 0 & -2 \\ 2 & 0 \end{matrix} \right).
\end{align}

Close to $\nu = 1/2$, we can approximate the composite electron Hall resistivity by its classical value,
\begin{align}
\label{comelectronHall}
\rho_{yx}^{\text{cf}} = \frac{1}{n_e e c} B_{\text{eff}} = \frac{h}{e^2} {1 - 2 \nu \over \nu},
\end{align}
where $B_{\text{eff}} = B - 2 \phi_0 n_e$ is the effective magnetic field seen by the composite electrons, $B$ is the total applied magnetic field, 
$n_e$ is the electron density, and $\nu = \phi_0 n_e/ B$. 
Deviations away from half-filling are parameterized using:
\begin{align}
\label{deviation}
\delta = 1 - 2 \nu = \frac{n_h - n_e}{n_e + n_h}. 
\end{align}
Note that $\delta \rightarrow -\delta$ under the particle-hole transformation, $\nu \rightarrow 1 - \nu$.

The longitudinal resistivity of the composite electrons can be taken to be
\begin{align}
\label{drude}
\rho_{xx}^{\text{cf}} = \rho_{yy}^{\text{cf}} = \frac{m_\psi}{n_e e^2 \tau^{\rm cf}_{\text{tr}}},
\end{align}
where $\tau^{\rm cf}_{\text{tr}}$ is the composite electron transport scattering time.
Eqn. (\ref{CFLcomposition}) tells us that the electrical longitudinal resistivity $\rho^{\rm CFL}_{xx} = \rho_{xx}^{\rm cf}$.

The electrical conductivity tensor is obtained by inverting $\rho^{\rm CFL}$:
\begin{align}
\sigma_{xy}^{\text{CFL}} = \frac{e^2}{h} \frac{2+{\delta \over \nu}}{(\bar{\rho}^{\rm cf}_{xx})^2 + (2+{\delta \over \nu})^2} ,
\end{align}
where $\bar{\rho}^{\rm cf}_{xx} = \rho^{\rm cf}_{xx} \frac{e^2}{h}$ is the longitudinal resistivity in units of $\frac{h}{e^2}$. 
Note that at exactly half-filling, when $\delta = 0$, we have:
\begin{align}
\label{cflsigmaxy}
\sigma_{xy}^{\text{CFL}} = \frac{e^2}{h} \frac{2}{(\bar{\rho}_{xx}^{\text{CFL}})^2 + 4} < \frac{1}{2} \frac{e^2}{h} ,
\end{align}
where $\bar{\rho}_{xx}^{\text{CFL}} = \bar{\rho}_{xx}^{\text{cf}}$ is the measured longitudinal resistivity at 
half-filling, in units of $h/e^2$ (in the CFL state). 
Thus, the Hall conductivity of the CFL, at exactly half-filling, is strictly less than $\frac{1}{2} \frac{e^2}{h} $.

\subsubsection{Anti-CFL}

Now let us consider the resistivity of the Anti-CFL state. 
As before, the Ioffe-Larkin rule allows us to write the resistivity of the holes as:
\begin{align}
\label{rhoH}
\rho^{\text{h}} &= \rho^{\text{ch}} - \rho^{\text{cs}},
\end{align}
where $\rho^{\text{ch}}$ is the resistivity of the composite holes and $- \rho^{\text{cs}}$ is the resistivity due to the attached flux tubes, which is given by Eqn. (\ref{rhoCS}). 
The minus sign in front of $\rho^{\rm cs}$ relative to the corresponding expression in Eqn. (\ref{CFLcomposition}) is due to the fact that the composite holes are attached to $+2$ units of statistical flux as opposed to the $-2$ units of flux which are coupled to the composite electrons.
Close to $\nu  =1/2$, we approximate the Hall resistivity of the composite holes as
\begin{align}
\rho_{yx}^{\text{ch}} = \frac{1}{n_h e c} B_{\text{eff}} = {h \over e^2} {1 - 2 \nu \over 1 - \nu} = {h \over e^2} {\delta \over 1 - \nu},
\end{align}
where we have used the relation $n_e + n_h = B/\phi_0$ and the fact that the composite holes feel the same effective magnetic field as the composite electrons.
Comparing to the composite electron Hall resistivity in Eqn. (\ref{comelectronHall}), we notice the replacement of $\nu^{-1}$ by $(1- \nu)^{-1}$.

We write the resistivity of the composite holes as
\begin{align}
\rho_{xx}^{\text{ch}} = \rho_{yy}^{\text{ch}} = \frac{m_\chi}{n_h e^2 \tau_{\text{tr}}^{\text{ch}}},
\end{align}
where $\tau_{\text{tr}}^{\text{ch}}$ is now the scattering time for the composite holes and $m_\chi$ is the renormalized composite hole mass.
Precisely at half-filling and in the limit where particle-hole symmetry is preserved (i.e. zero Landau level mixing), the composite holes
and composite electrons should have the same diffusion constant, so that $\sigma_{xx}^{\text{cf}} = \sigma_{xx}^{\text{ch}}$ at half-filling. 
Since the off-diagonal components of $\sigma^{\text{ch}}$ and $\sigma^{\text{cf}}$ vanish at $\nu =1/2$, this also implies
\begin{align}
\rho_{xx}^{\rm ch} = \rho_{xx}^{\rm cf} \;\; \text{ at } \nu=1/2.
\end{align} 
In general, since the composite hole density is magnetic field dependent, $\tau_{\text{tr}}^{\text{ch}}$
will be different from $\tau_{\text{tr}}^{\text{cf}}$ in its field dependence as we move away from half-filling.

The electrical conductivity of the full system in the Anti-CFL phase is given in terms of the conductivity of the holes and the single filled Landau level:
\begin{align}
\label{sigmaACFL}
\sigma^{\text{ACFL}} = (\rho^{\text{h}})^{-1} + \frac{e^2}{h}\left( \begin{matrix} 0 & 1 \\ -1 & 0 \end{matrix} \right).
\end{align}
Setting $\delta = 0$ in order to sit precisely at half-filling, we find
\begin{align}
\label{acflsigmaxy}
\sigma_{xy}^{\text{ACFL}} = \frac{2 + (\bar{\rho}_{xx}^{\text{ch}})^2}{4 + (\bar{\rho}_{xx}^{\text{ch}})^2}  > \frac{1}{2} \frac{e^2}{h},
\end{align}
where $\bar{\rho}_{xx}^{\rm ch} = {e^2 \over h} \rho_{xx}^{\rm ch}$. Using Eqns. (\ref{rhoH}), (\ref{sigmaACFL}), the composite hole resistivity can be 
written in terms of the physical, measured conductivity as 
\begin{align}
\rho_{xx}^{\text{ch}} = \rho_{xx}^{\text{h}} = \frac{\sigma_{xx}^{\text{ACFL}}}{(\sigma_{xx}^{\text{ACFL}})^2 + (\sigma_{xy}^{\text{ACFL}} - \frac{e^2}{h})^2}
\end{align}
Remarkably, the Hall conductivity of the Anti-CFL, at exactly half-filling, is strictly greater than $\frac{1}{2} \frac{e^2}{h} $.

From the expressions for the Hall conductivities at half-filling ($\delta = 0$) computed within the CFL and Anti-CFL theories, 
Eqns. (\ref{cflsigmaxy}) and (\ref{acflsigmaxy}), we observe the ``sum rule":
\begin{align}
\sigma_{xy}^{{\rm CFL}} + \sigma_{xy}^{{\rm ACFL}} = {e^2 \over h}.
\end{align}
At the critical point presented in Sec. \ref{trans}, we find the Hall conductivity to equal ${1 \over 2} {e^2 \over h}$ within a linear response treatment. 

\subsubsection{Comparison of the CFL and Anti-CFL Resistivity}

Therefore, in comparing the resistivity tensors for CFL and Anti-CFL, at $\nu = 1/2$, we find:
\begin{align}
\rho_{yx}^{\text{CFL}} &= 2 \frac{h}{e^2} & \rho_{yx}^{\text{ACFL}} &= \frac{h}{e^2} \frac{2 + (\bar{\rho}_{xx}^{\text{ch}})^2}{1 + (\bar{\rho}_{xx}^{\text{ch}})^2}
\nonumber \\
\rho_{xx}^{\text{CFL}} &= \rho_{xx}^{\text{cf}} & \rho_{xx}^{\text{ACFL}} &= \frac{\bar{\rho}_{xx}^{\text{ch}}}{1 + (\bar{\rho}_{xx}^{\text{ch}})^2} ,
\end{align}
where the composite particle resistivities can be obtained in terms of physical, measurable quantities as described in the previous subsections. 

If the system is uniformly in the CFL state for $\nu < 1/2$ and in the Anti-CFL state for $\nu > 1/2$, the resulting resistivity will be given by:
\begin{align}
\label{resistivitypiece}
\rho = \left\{ \begin{array}{ccc}
\rho^{\text{CFL}} & \text{ if } & \delta > 0 \;\; (\nu < 1/2)\\
\rho^{\text{ACFL}} & \text{ if } & \delta < 0 \;\; (\nu > 1/2).\\
\end{array} \right.
\end{align}
This piece-wise behavior implies a small jump in the electrical resistivity tensor at $\nu = 1/2$:
\begin{align}
\label{jumps}
\Delta \rho_{xx} \equiv & \left.\rho_{xx}\right|_{\delta = 0^-} - \left.\rho_{xx}\right|_{\delta = 0^+} 
= -\frac{h}{e^2} (\bar{\rho}_{xx}^{(-)})^3, \cr
\Delta \rho_{yx} \equiv & \left.\rho_{yx}\right|_{\delta = 0^-} - \left.\rho_{yx}\right|_{\delta = 0^+} = -\frac{h}{e^2} (\bar{\rho}_{xx}^{(-)})^2. 
\end{align}
In Eqn. (\ref{jumps}), we have denoted the jumps in terms of 
\begin{align}
\bar{\rho}_{xx}^{(-)} \equiv \bar{\rho}_{xx}(\delta = 0^{-})
\end{align}
which is equal to the resistivity at half-filling as approached from $\nu < 1/2$.
Since the jumps are proportional to a power $\bar{\rho}_{xx}^{(-)}$, they would be more noticeable for somewhat more disordered samples. 

In experiment,\cite{willett1997} $\rho_{xx}$ at $\nu = 1/2$ is 
typically observed to range from $100 \Omega/\square$ to $5$ k$\Omega/\square$. 
Taking the upper limit, $\bar{\rho}^{(-)} = .19$, we find
that the jumps, $\Delta \rho_{xx} \approx - 200 \Omega/\square$ and $\Delta \rho_{yx} \approx - 1$ k$\Omega/\square$.
These jumps correspond to roughly $4\%$ and $2\%$ differentials, $\Delta \rho_{ij} / \rho_{ij}$.

In addition to the jump in electrical resistivity, there are also discontinuities in the slopes, 
$\frac{d \rho_{xx}}{d \delta}$ and $\frac{d \rho_{yx}}{d \delta}$. 
To compute these discontinuities, we need to know the behavior of $\rho_{xx}^{\rm cf}$ and $\rho_{xx}^{\rm ch}$ as one moves away from half-filling.
Let us assume the following analytic expansions:
\begin{align}
\rho_{xx}^{\text{cf}} = & \bar{\rho}_{xx}^{(-)}\Big(1 + \sum_{n=1}^\infty c_n \delta^n\Big), \cr
\rho_{xx}^{\text{ch}} = & \bar{\rho}_{xx}^{(-)}\Big(1 + \sum_{n=1}^\infty c'_n \delta^n\Big),
\end{align}
where the $c_n$ and $c'_n$ are model-dependent constants.
Performing an expansion of the longitudinal and Hall resistivities first in powers of 
$\bar{\rho}_{xx}^{(-)}$ and then in the deviation $\delta$ away from half-filling, 
we find slope discontinuities:
\begin{align}
\label{slopedis}
\Delta \bar{\rho}'_{xx} \equiv \left. \frac{d \bar{\rho}_{xx}}{d \delta} \right|_{\delta = 0^-} - \left. \frac{d \bar{\rho}_{xx}}{d \delta} \right|_{\delta = 0^+} = & \bar{\rho}_{xx}^{(-)} \Big(4 + c'_1 - c_1\Big) \cr
\Delta \bar{\rho}'_{yx} \equiv \left. \frac{d \bar{\rho}_{yx}}{d \delta} \right|_{\delta = 0^-} - \left. \frac{d \bar{\rho}_{yx}}{d \delta} \right|_{\delta = 0^+} = & - 2 (\bar{\rho}_{xx}^{(-)})^2 \Big(3 + c'_1\Big).
\end{align}
If either $c_1$ or $c'_1$ are non-zero, along with any higher-order coefficients in the assumed analytic expansion, these slope discontinuities -- if observed -- could provide constraints on theoretical models.

To clarify the above expressions, it is helpful to consider two typical ways of departing away from half-filling: (i) we may fix the electron density $n_e$ and vary the magnetic field $B$, or (ii) we may fix $B$ and vary $n_e$.
In the former case, we may write $\delta = 1 - (1 + (B - B_{1/2})/B_{1/2})^{-1} \sim (B - B_{1/2})/B_{1/2}$, where $B_{1/2} = 2 \phi_0 n_e$.
In the latter case, we may write $\delta = - (n_e - (n_e)_{1/2})/(n_e)_{1/2}$, where $(n_e)_{1/2} = B/(2 \phi_0)$.

To get a feeling for possible values of $c_1$ and $c'_1$, we can estimate\cite{halperin1993} $\rho_{xx}^{\rm cf}$ using Eqn. (\ref{drude}), with
\begin{align}
\label{scatteringrate}
(\tau_{\rm tr}^{\rm cf})^{-1} = {4 \pi n_{\rm imp} \over m_\psi k^{\rm cf}_F d_s},
\end{align}
where the impurity density $n_{\rm imp}$ is contributed by a doping layer a distance $d_s$ away from the electron gas and $k^{\rm cf}_F$ is the Fermi momentum of the composite electrons.
A similar estimate can be given for $\rho^{\rm ch}$. 
Note, however, that $n_{\rm imp} = n_e$ for {\it both} $\tau_{\rm tr}^{\rm cf}$ and $\tau_{\rm tr}^{\rm ch}$, which directly leads to the absence of a symmetry relating the expansion parameters $c_m$ to $c'_n$ away from half-filling.
We find $\rho_{xx}^{\rm cf} \propto (n_e)^{-1/2}$ and $\rho_{xx}^{\rm ch} \propto (n_e)^{-1/2}(1+{2 \delta \over 1 - \delta})^{-3/2}$.
Fixing $n_e$ and varying $B$ away from half-filling, we find $c_1 = 0$ and $c'_1= - 3$.
Using again $\bar{\rho}_{xx}^{(-)} = .19$, we find the slope discontinuities: $\Delta \bar{\rho}'_{xx} = .19$ and $\Delta \bar{\rho}'_{yx} = 0$ to ${\cal O}((\bar{\rho}_{xx}^{(-)})^2)$.
Within this model, the first contribution to the discontinuity in the slope of the Hall resistivity occurs at quartic order, $\Delta \bar{\rho}'_{yx} = 2 (\bar{\rho}_{xx}^{(-)})^4 (5 + 2 c_1') = - .003$.

Therefore, if the system is uniformly in the CFL state for $\nu < 1/2$, and uniformly in the Anti-CFL state for $\nu  >1/2$, 
we expect the above discontinuities in the resistivity given in Eqns. (\ref{jumps}) and Eqn. (\ref{slopedis}). 

Experimental observation of the jumps implies an interesting interplay of disorder and strong interactions since the very existence of the CFL requires the disorder to not be too strong relative to the strength of the electron-electron interactions. For sufficiently strong disorder, which smears out the incompressible FQH states,
the system at half-filling is presumably perched at the integer quantum Hall plateau transition.

Deviations from the above predictions are expected if, rather than a sharp first order transition 
between the two states, the transition is continuous. As discussed in Sec. \ref{trans}, the transition could be continuous in the absence of quenched disorder, or it could be intrinsically first order, but rounded by the effects of disorder.

\subsection{Magnetoresistance Oscillations and the Princeton Experiment }
\label{expPrinceton}

A successful method of detecting the Fermi sea of composite fermions in the CFL is by imposing a periodic potential on the system,
modulated in one direction with a wave vector $q \ll k_F$. As the magnetic field is varied, the system exhibits oscillations in the 
magnetoresistance, which are determined by when the diameter of the cyclotron motion is commensurate with the period $\lambda = 2\pi/q$. 
These oscillations occur for magnetic field deviations away from half-filling that are smaller than conventional Shubnikov-de Haas oscillations.

For charge $-e < 0$ fermions in a circular Fermi surface with Fermi wave vector $k_F$ in a uniform magnetic field $B$, the cyclotron radius $R_c$ is given by
\begin{align}
R_c = \frac{\hbar k_F}{e B} . 
\end{align}
Consider a modulation of the electric potential and the magnetic field, such that
\begin{align}
U(x) &= U_q \cos(q x) 
\nonumber \\
B(x) &= B_0 +  B_q \cos(q x). 
\end{align}
Let $\tau$ be the transport scattering time of the fermions, and $\omega_c = \frac{eB_0}{mc}$. To leading order in $(\tau \omega_c)^{-1}$, and for weak 
modulations in the fields, the change in the resistivity is\cite{gerhardts1996}
\begin{widetext}
\begin{align}
\label{deltaRho}
\Delta \rho_{xx} = \rho_0 \frac{2 \tau^2}{h^2 n} \frac{q}{R_c}\left( \frac{B_0}{|B_0|} U_q \cos(q R_c - \pi/4) + \frac{k_F}{q} \frac{\hbar e B_q}{mc} \sin(qR_c - \pi/4)\right)^2 .
\end{align}
\end{widetext}
Here $\rho_0$ is the longitudinal resistivity without the modulation, $\tau$ is the transport scattering time, $n$ is the density of fermions, $m$ is their mass,
and the cylotron radius $R_c$ is defined relative to the uniform part of the magnetic field, $B_0$. 
Eq. (\ref{deltaRho}) is obtained from a semiclassical calculation, and is therefore valid for temperatures $T_c < T < \frac{\pi k_F}{q} T_c$, where
$T_c = \frac{\hbar \omega_c}{k_B 2\pi^2}$. As discussed in Ref. \onlinecite{gerhardts1996}, this condition is so that the temperature is sufficiently high so that the sum over Landau levels
can be converted to an integral over energies, but low enough that only states near the Fermi surface are excited. Outside of this range, a fully
quantum mechanical calculation is required.\cite{peeters1993}

For the CFL, the modulation in electric potential, $U(x)$, induces a density modulation
\begin{align}
n_e(x) = n_{e0} + K_{00} U_q \cos(q x),
\end{align}
where $n_{e0}$ is the electron density in the absence of the potential modulation, and
$K_{00}$ is the compressibility of the CFL state. 
This was estimated in RPA to be\cite{halperin1993}
\begin{align}
K_{00}(\omega = 0, q) \approx \frac{m_\psi}{\hbar^2 (m_\psi V(q) + 10\pi/3)},
\end{align}
where $V(q)$ is the Fourier transform of the interaction potential, which for short-ranged interactions vanishes as $q \rightarrow 0$. 

The density modulation in turn induces a modulation in the effective magnetic field felt by the composite electrons of the CFL,
\begin{align}
B_{\text{eff}} = (B - B_{1/2}) - 2 \phi_0 K_{00} U_q \cos(q x) ,
\end{align}
where $B_{1/2} = 2 \phi_0 n_{e0}$ is the magnetic field at exactly half-filling. 
Eq. (\ref{deltaRho}) indicates that the ratio of the amplitudes from the potential modulation and the magnetic field modulation is 
$\frac{\pi m_\psi}{h^2 K_{00}} \frac{l_B}{\lambda} \approx \frac{l_B}{\lambda}$ for the CFL with short-ranged interactions.
Typically, $\lambda \sim 500$ nm while $l_B \sim 10$ nm, indicating that the 
modulation from the potential term can be ignored; including it yields a correction to the resistivity of only $\sim 1 - 2$\%. 

Since $\rho_{ii} = \rho_{ii}^{\text{cf}}$ in the CFL, the only component of the resistivity that receives a noticeable modulation is 
$\rho_{xx}$ (at higher orders, all components obtain an oscillatory component), and is given by eq. (\ref{deltaRho}). Minima in the longitudinal resistance $\rho_{xx}$ therefore occur when
\begin{align}
2 R_c = \lambda ( l + 1/4), 
\end{align}
for integer $l$. In the CFL, the composite fermion density is equal to the electron density, so that $k_F = \sqrt{4 \pi n_{e0}} = \sqrt{2\pi B_{1/2}/\phi_0}$. 
Moreover, the effective magnetic field seen by the composite fermions is $B_{\text{eff}} = \delta B = B - B_{1/2}$, 
so that the resistance minima are expected to occur whenever
\begin{align}
\label{cflMin}
\delta B   = \frac{2 \hbar \sqrt{4\pi n_{e0}}}{e \lambda (l + 1/4)} .  
\end{align}

In the case of the Anti-CFL, the composite holes couple with opposite charge to the external electromagnetic field. Therefore they would effectively feel a potential
modulation with opposite sign:
\begin{align}
U_{\text{eff}}(x) = -U_q \cos(q x). 
\end{align}
This induces a change in the hole density:
\begin{align}
n_h(x) = n_{h0} - K_{00} U_q \cos(q x)) . 
\end{align}
The effective magnetic field felt by the composite holes is 
\begin{align}
B_{\text{eff}} &= (B - B_{1/2}) - 2 \phi_0 K_{00} U_q \cos(q x)) 
\end{align}
This then leads to a modulated resistivity for the composite holes and therefore also the holes:
\begin{widetext}
\begin{align}
\label{deltaRhoxxacfl}
\Delta \rho_{xx}^{\text{h}} = \Delta \rho_{xx}^{\text{ch}} = \rho_0^{\text{ch}} \frac{2 (\tau_{\text{tr}}^{\text{ch}})^2}{h^2 n_h} \frac{q}{R_c}\left( -\frac{B_0}{|B_0|} U_q \cos(q R_c - \pi/4) + \frac{k_F}{q} \frac{\hbar e B_q}{m_\chi c} \sin(qR_c - \pi/4)\right)^2 ,
\end{align}
\end{widetext}
with $B_q = -2 \phi_0 K_{00} U_q $ and $B_0 = B - B_{1/2}$, while the other components of $\rho^{\text{ch}}$ remain unmodulated. 
Note that in this case, the Fermi wave vector of the composite holes is  $k_F = \sqrt{4\pi n_{h0}} = \sqrt{4\pi (\nu^{-1} - 1) n_{e0}}$. 
The resistivity tensor of the electron system in terms of the composite hole resistivity tensor is more complicated than the analogous problem in 
the case of the CFL:
\begin{align}
\rho^{ACFL} = [ (\rho^\text{h})^{-1} + \frac{e^2}{h} i \sigma^y]^{-1},
\end{align}
where $\sigma^y = \left(\begin{matrix} 0 & - i \\ i & 0 \end{matrix} \right)$. If we expand the above equation to lowest order in $\rho_{xx;0}^h$, we find:
\begin{align}
\Delta \rho^{ACFL}_{xx} = \frac{\rho_{0}^{ACFL}}{\rho_0^{\text{h}}} \Delta \rho_{xx}^\text{h}+ \mathcal{O}((\rho_{0}^h)^3 ) .
\end{align}
Therefore, in the Anti-CFL, ignoring once again the contribution from the potential modulations,
one expects resistance minima to occur whenever $\delta B$ satisfies the following equation: 
\begin{align}
\label{acflMin}
\delta B =  \frac{2 \hbar \sqrt{4\pi n_{h0}}}{e \lambda (l + 1/4)},
\end{align}
for integer $l$. 

In the experiment of Kamburov et. al., Ref. \onlinecite{kamburov2014}, it was observed that the resistance minima away from $\nu = 1/2$ satisfied
Eqn. (\ref{cflMin}) for $\nu < 1/2$, whereas the resistance minima are much better approximated by Eqn. (\ref{acflMin}) for $\nu  > 1/2$. In other words, 
the experiment is explained by considering the system to be in the CFL state when $\nu < 1/2$, and in the Anti-CFL state for $\nu > 1/2$. 

We note that in the CFL, the potential modulation and effective magnetic field modulation are exactly $\pi$ out of phase relative to each other. This is due to the fact that in the CFL, $-2$ units of flux are attached to each composite fermion. However in the Anti-CFL, we see that the effective potential modulation and effective magnetic field modulation, $U_{\text{eff}}$ and $B_{\text{eff}}$, are exactly in phase. From eq. (\ref{deltaRhoxxacfl}), we see that the phase difference between the potential and magnetic field modulation enters the formula for the modulation in $\Delta \rho_{xx}$. Even though the effect of the potential modulation is weak compared to that of the effective magnetic field modulation, it would be interesting to see if the effect of these relative phases can be extracted experimentally. 

The analysis in Ref. \onlinecite{gerhardts1996} and its adaptation to the CFL and Anti-CFL reviewed above for 1D periodic potentials also applies to 2D periodic potentials. 
Thus, we expect minima of the magnetoresistance oscillations in such circumstances to reflect the Fermi wave vector of the underlying state, consistent with recent observations.\cite{LiuDengWaltz}

\subsection{Magnetoresistance Oscillations Across CFL - Anti-CFL Interfaces}
\label{expTrans}

An important consequence of the theory that we have developed here is the fact that the CFL - Anti-CFL interface possesses a ``partially fused'' interface phase where the
composite electrons in the CFL can directly tunnel into composite holes of the Anti-CFL. As emphasized above, this is a highly non-trivial phenomena, because the composite
electrons of the CFL and the composite holes of the Anti-CFL represent \it topologically distinct \rm degrees of freedom of the electron fluid. Here we would like 
to investigate methods to detect such a phenomenon. 

\begin{figure}
	\centering
	\includegraphics[width=3.2in]{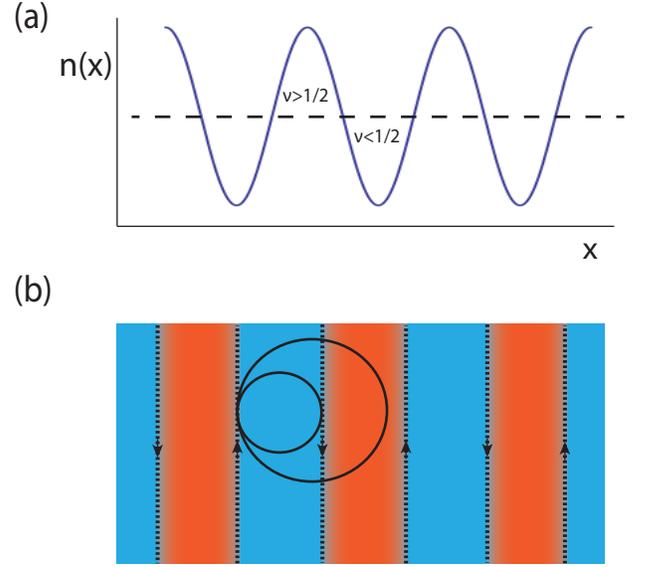}
	\caption{\label{cflStripes} (a) Density modulation of the 2DEG, induced by a periodic potential. The dashed line represents the density at exactly $\nu = 1/2$. Regions above the dashed line have
$\nu > 1/2$ and are in the Anti-CFL phase, while regions below the dashed line have $\nu < 1/2$ and are in the CFL phase. (b) The system is thus broken up into alternating strips of CFL 
(blue) and Anti-CFL (red). The gapless chiral boson interface fields are robustly present at the interfaces, which we have assumed to be in ``partially fused interface 2'' of Sec. \ref{cfl-acfl}. 
Observation of magnetoresistance oscillations would indicate the ability for composite electrons
in the CFL to directly tunnel into the composite holes of the Anti-CFL. The circles indicate the effective cyclotron orbits of the composite particles that are commensurate with the periodic modulation. 
As the magnetic field is tuned, the dashed line in (a) moves vertically relative to the density modulation, thus changing the relative
widths of the CFL and Anti-CFL regions. 
}
\end{figure}

To this end, let us consider the application of a periodic potential to the electron fluid, 
\begin{align}
U(x) = U_q \cos(q x),
\end{align}
where $q$ is the wavelength of the potential modulation. This induces a density modulation in the electron fluid:
\begin{align}
n_e(x) = n_{e0} + K_{00} U_q \cos(q x).
\end{align}
We would like to consider the amplitude $U_q$ to be large enough so that this potential induces striped domains of alternating 
regions with $\nu > 1/2$ and $\nu < 1/2$, which would induce alternating regions of CFL and Anti-CFL (see Fig. \ref{cflStripes}). 
Since the electron tunneling across the interface between $\nu < 1/2$ and $\nu  > 1/2$ is as strong as it is in the rest of the system, we expect that the CFL - Anti-CFL interfaces would be in a partially fused phase (see Sec. \ref{cfl-acfl}). 

Next, consider the magnetoresistance $\rho_{xx}(B)$, as a function of the magnetic field $B$ tuned away from half-filling. The composite electrons of the CFL directly tunnel into the
composite holes of the Anti-CFL, so one expects magnetoresistance oscillations $\rho_{xx}(B)$, with a period set by the Fermi wave vector of the composite particles and the wavelength
of the modulation. As $B$ is tuned, the width of the CFL regions will change relative to the Anti-CFL regions, so we would 
like to consider the case that the modulation $U_q$ is large enough so that over the range
where $B$ is tuned, the system always consists of alternating regions of CFL and Anti-CFL. The existence of the magnetoresistance oscillations in this regime would be non-trivial verification of
the fact that composite electrons of the CFL can directly tunnel into composite holes of the Anti-CFL. 

\subsection{Effect of Landau level mixing}

It is important to consider the effects of Landau level mixing, which explicitly breaks particle-hole symmetry, 
on interactions within the lowest Landau level. The effects of Landau level mixing can be quantified using 
the parameter, $\kappa = (e^2/4\pi \epsilon l_B)/\hbar \omega_c$, where $\epsilon$ is 
the dielectric constant, the magnetic length $l_B = \sqrt{\hbar c/e B}$, and $\omega_c = e B/m_e c$ is the cyclotron frequency of the electron.
In GaAs heterostructures, $\kappa \sim .6 -1.8$ for magnetic fields $B = 2 - 15$ T.
Interestingly, it has been found that corrections to the effective interactions in the lowest Landau level 
due to Landau level mixing are numerically small even though the naive expansion parameter, $\kappa \sim {\cal O}(1)$.\cite{BisharaNayak}

Since non-zero $\kappa$ explicitly breaks particle-hole symmetry, it is expected that weak Landau level mixing will shift 
the location of the transition between the CFL and Anti-CFL away from $\nu=1/2$. Strong Landau level mixing, or large 
$\kappa$, could potentially eliminate the transition altogether, leaving either the CFL or Anti-CFL dominant around half-filling. 
Therefore, a qualitative result that would validate the theory presented here would be to observe that systems with weak
Landau level mixing have the composite fermion density equal to the electron density for $\nu < 1/2$ and the hole density for 
$\nu > 1/2$, while other systems with much stronger Landau level mixing have the composite fermion density equal to the electron
density (or hole density) on both sides of $\nu = 1/2$. The precise quantitative value of Landau level mixing $\kappa$ 
needed to see the latter case is a question that requires further detailed study.

\section{Discussion and Conclusion}

In this paper, we have introduced the particle-hole conjugate of the CFL, which we refer to as the Anti-CFL.
In the presence of particle-hole symmetry, we suggest that the Anti-CFL plays an equally important role 
as the CFL in determining the low energy physics around the half-filled Landau level. 
We posit that the CFL is operative for fillings fractions $\nu<1/2$ and the Anti-CFL is favored for fillings $\nu>1/2$.

As outlined in Sec. \ref{expPrinceton}, the existence of the Anti-CFL for $\nu > 1/2$  naturally explains the remarkable experiment of Kamburov et 
al.\cite{kamburov2014} measuring magnetoresistance oscillations about half-filling in the presence of a 1D periodic potential.
The locations of the magnetoresistance minima reveal the value of the Fermi wave vector of the underlying composite particles of the CFL or Anti-CFL.

A variety of other experiments that probe the Fermi wave vector of the composite particles should be revisited to see whether 
they are consistent with the predictions of CFL versus Anti-CFL: for instance, magnetoresistance oscillations in the 
presence of a 2D anti-dot superlattice\cite{kang1993}, magnetic focusing\cite{GoldmanSuJainfocusing, SmetWeissFocusing, SmetFleischmannFocusing}, and composite particle cyclotron resonances\cite{KukushkinSmetCyclotron}.

A further consequence of this picture is borne out in the behavior of the DC resistivity as the filling fraction is varied across half-filling.
Assuming a first order transition, we find that the resistivity tensor exhibits both a jump discontinuity and slope change as the state 
transforms from the CFL into the Anti-CFL across half-filling. Observation of this prediction would be an independent verification 
of the theory presented here. A second order transition, on the other hand, would not be expected to yield a jump discontinuity in the resistivity. 

In addition, we have developed a local field theory for both the CFL and Anti-CFL, valid when the system has a boundary.
This theory enables us to study various interfaces between these two states. We suggest the existence of interface phases 
which are characterized by intermediate, partially fused phases in addition to either decoupled or, in the case of CFL - I -CFL junctions, fully healed interfaces.
These phases are characterized by the ability of the composite electrons or holes to tunnel into one another, even while the interface fails to fully heal.

We have also put forward a theory that is capable of describing a continuous transition between the CFL and Anti-CFL.
Interestingly, the transition point exhibits non-Fermi liquid behavior reminiscent of the marginal Fermi liquid, independent of whether 
or not the Coulomb interaction is short-ranged.

There are a number of interesting directions for further research. The gauge-invariant boundary theory of 
the CFL and Anti-CFL described in Sec. \ref{edgeTheories} should enable the computation of scaling dimensions of operators inserted along the boundary. 
The scaling dimension of the electron operator, in particular, would allow the computation of the tunneling current into the edge from a nearby metallic lead. 
Such a calculation would make contact with previous work in Ref. \onlinecite{shytov1998, levitov2001} and possible experimental measurements. 

The boundary theory in Sec. \ref{edgeTheories} and existence of a fully fused CFL-I-CFL interface phase described in Sec. \ref{interfaces} should also 
furnish a complementary approach to the calculation of the bulk electronic tunneling density of states.\cite{he1993, kim1994}
In the mean-field approximation for computing the electron correlation functions, Eqn. (\ref{mfCorr}), 
it is straightforward to see that the electron 
auto-correlation function at the fully fused interface decays with imaginary time $\tau$ as $\frac{1}{\tau}\exp(- m \tau)$. 
Laplace transforming gives Eqn. (\ref{bulkTDOS}) for $\eta  = 2$. 
We expect a RPA treatment of the gauge field to be able to recover the formula of Eqn.  (\ref{bulkTDOS}) for other values of $\eta$ as well. 

In Sec. \ref{interfaces}, we provided arguments for the existence of various line interface phases.
A detailed calculation for the different regimes in parameter space where such phases occur would be valuable.

It would be extremely interesting to consider energetic and entanglement properties of model wave functions for the CFL 
and Anti-CFL described in Sec. \ref{antiCFLbulk} and to compare with numerical experiments studying the ground state of interacting electrons in the lowest Landau level.
Such numerical experiments might provide insight into the underlying reason why the CFL might be favored for filling fractions $\nu < 1/2$, while the Anti-CFL obtains for $\nu>1/2$.

At a more qualitative level, it is interesting to consider additional implications of the Anti-CFL state for physics about half-filling. 
For example, strong electron-electron interactions and weak disorder are crucial to the stabilization of the CFL and Anti-CFL physics
at half-filling. As the electron-electron interaction is weakened or disorder is increased, it is expected that the system
will crossover into a regime which corresponds to the regime of the integer quantum Hall plateau transition. 
A theoretical understanding of the nature of this interpolation is an open problem and consideration of the Anti-CFL
makes any such interpolation even more interesting.

\section{Acknowledgments}

It is a pleasure to thank Jason Alicea, Eduardo Fradkin, Steve Kivelson, Chetan Nayak, Sri Raghu, Javad Shabani, and Steve Simon for helpful discussions. This research was supported in part by the National Science Foundation, under Grants No.  DMR-14-04230 (M.P.A. Fisher), by the 
Caltech Institute of Quantum Information and Matter, an NSF Physics Frontiers Center with support of the Gordon and Betty 
Moore Foundation (M.P.A.F.) and by the Gordon and Betty Moore Foundation’s EPiQS Initiative through Grant GBMF4304 (M.P.A. Fisher).
M.M. is partially supported by a grant from the Templeton Foundation on Emergence and Entanglement.
M.M. acknowledges the support and hospitality of Microsoft Station Q, where this work was initiated, and the Kavli Institute for 
Theoretical Physics at the University of California Santa Barbara, where this work was completed.
Upon completion of this manuscript, we became aware of the recent preprint \cite{Soncompositefermion} studying particle-hole symmetry in the half-filled 
Landau level from a different perspective. We thank D. T. Son for sharing a draft of his paper with us before publication.

\appendix

\section{General relation between particle-hole conjugation and vortex duality}
\label{dualitySec}

For every fermionic FQH state at filling fraction $\nu$, one can consider the ``particle-hole'' conjugate of that FQH state, which occurs when the holes are at a filling fraction $\nu$ and the electrons are at filling $1 - \nu$. Bosonic FQH states on the other hand do not have a natural notion of the particle-hole conjugate state. For them, the natural analog is to consider the vortex dual of a given bosonic FQH state. Below we will first give a general discussion of vortex duals of bosonic FQH states, and subsequently illustrate the intimate relation between particle-hole duals of fermionic FQH states and vortex duals of bosonic FQH states. 

\subsection{Vortex dual of bosonic FQH states}

Given a system of bosons, one can always choose to describe the system in terms of the dynamics of its vortices. When the bosons are at a filling fraction $\nu = p/q$, this means there are $p$ bosons per $q$ units of flux quanta, or $p$ bosons per $q$ vortices. Since the vortices see the original particles as sources of magnetic flux, this means that there are $q$ vortices for every $p$ units of their effective flux quanta. In fact it is easy to see that the vortices are at an effective filling fraction $\nu_v = - \nu^{-1} = -q/p$, as we will show in more detail below. 

A general effective theory for the vortices takes the form
\begin{align}
\mathcal{L} = \frac{1}{2\pi} A^E \partial a + \mathcal{L}_v(a,\phi_v),
\end{align}
where $A_E$ is an external probe gauge field that couples to the particle current, and $\mathcal{L}_v(a, \phi_v)$ describes the state of the vortices. Observe that if state of the vortices results in a polarization tensor $\Pi_v$ for the gauge field $a$, then after integrating out $\phi_v$ we would obtain the effective action
\begin{align}
\mathcal{L} = \frac{1}{2\pi} A^E \partial a + \frac{1}{2} a_\mu \Pi_{v;\mu\nu} a_\nu. 
\end{align}
Integrating out $a$ gives the polarization tensor of the system:
\begin{align}
\mathcal{L} = \frac{1}{2} A^E_{\mu} \Pi_{\mu\nu} A^E_{\nu} .
\end{align}

It is easy to see that if the bosons are described by a conductivity tensor $\sigma$, then the vortices should be in a collective 
state with a conductivity tensor $\sigma_v = \sigma^{-1}$. In a clean system with continuum translation invariance, the 
conductivity is $\sigma = \left(\begin{matrix} 0 & \sigma_{xy} \\ -\sigma_{xy} & 0 \end{matrix} \right)$, with $\sigma_{xy} = 2\pi \nu$, 
in units where the charge of the boson and $\hbar$ are set to one.  Therefore, the 
vortices would have a conductivity $\sigma_v = \left(\begin{matrix} 0 & -\sigma^{-1}_{xy} \\ \sigma_{xy}^{-1} & 0 \end{matrix} \right)$. 
From this, we can conclude that if the original bosons are in a filling fraction $\nu$, then the vortices must be at a filling fraction $\nu_v = -1/\nu$. 

Therefore, for any given bosonic FQH state at filling fraction $\nu$, one can consider a dual 
state at filling fraction $-1/\nu$, where the vortices form that filling fraction $\nu$ FQH state. 

Let us consider several examples. First let us consider a class of incompressible FQH states, described by a $N \times N$ symmetric
integer matrix $K$. The effective theory is:
\begin{align}
\mathcal{L} = \frac{1}{4\pi} K_{IJ} a_I \partial a_J + \frac{1}{2\pi} q_I A^E \partial a_I,
\end{align}
where $q$ is the ``charge vector'' of the FQH states. Such a theory will describe a bosonic FQH state as 
long as the diagonal entries of $K$ are all even. 

The vortex dual of this state would therefore be described by
\begin{align}
\mathcal{L}_{\text{dual}} = \frac{1}{2\pi} A_E \partial a - \frac{1}{4\pi} K_{IJ} a_I \partial a_J + \frac{1}{2\pi} q_I a \partial a_I
\end{align}
In other words, it would be described by a new $(N+1) \times (N+1)$ matrix
\begin{align}
K_{v} = \left(\begin{matrix} 
-K & q \\
q^T & 0
\end{matrix} \right),
\end{align}
with a new charge vector $(q_v)_I = \delta_{I, N+1}$. 

As another example, note that there are two types of CFL-like states for bosons that we can consider at filling fraction $\nu =1$. One of them
is the usual CFL state of bosons, which is described by the effective action
\begin{align}
\mathcal{L}_{\text{bCFL}} = &\frac{1}{4\pi} a \partial a + \psi^\dagger (i \partial_t +a_t+ A_{Et} + \mu) \psi 
\nonumber \\
&+\frac{1}{2m_\psi}\psi^\dagger (i \partial_i +a_i + A_{Ei})^2 \psi 
\nonumber \\
&+\int d^2r' V(r-r') \psi^\dagger \psi (r) \psi^\dagger \psi (r').
\end{align}

Alternatively, we could consider a state where the vortices form an analogous $\nu_v = -1$ CFL state. The resulting effective theory would be given by
\begin{align}
\mathcal{L}_{\text{vCFL}} = & \frac{1}{2\pi} A_E \partial \tilde{a}
- \frac{1}{4\pi} \tilde{a} \partial \tilde{a} +\psi^\dagger (i \partial_t +a_t+ \tilde{a}_t + \mu) \psi 
\nonumber \\
&+\frac{1}{2m_\psi}\psi^\dagger (i \partial_i +a_i + \tilde{a}_i)^2 \psi 
\nonumber \\
&+\int d^2r' V(r-r') \psi^\dagger \psi (r) \psi^\dagger \psi (r').
\end{align}

$\mathcal{L}_{\text{bCFL}}$ and $\mathcal{L}_{\text{vCFL}}$ describe two thermodynamically distinct states of matter at filling fraction $\nu = 1$, both of which exhibit an emergent Fermi surface of composite fermions. $\mathcal{L}_{\text{vCFL}}$ was discussed previously in Ref. \onlinecite{alicea2005} as an alternative theory for the $\nu = 1$ CFL state of bosons. The logarithmic interaction between the vortices, and consequently the composite fermions in $\mathcal{L}_{\text{vCFL}}$, suppresses the gauge fluctuations of $a$. This leads to a marginal Fermi liquid state for the composite fermions, where the composite fermion self-energy only acquires logarithmic corrections from the gauge fluctuations. This theory is presumably equivalent to the theory developed in Ref. \onlinecite{read1998}, for the CFL at $\nu = 1$. 

\subsection{Fermionic particle-hole conjugates}

Here we would like to explain how particle-hole conjugates of fermion FQH states can generally be understood in terms of the vortex duals of bosonic FQH states discussed in the previous section. To see this generally, recall that in Sec. \ref{antiCFLbulk} we showed that if we think of the electron as
\begin{align}
c = \tilde{b} f,
\end{align}
where $f$ is a fermion that forms a $\nu_f = 1$ IQH state and $\tilde{b}$ is a ``composite boson.'' The effect of the $f$ fermion is effectively to attach one unit of flux to $\tilde{b}$, converting it into the electron $c$. In terms of these variables, the hole is described by
\begin{align}
h = f^\dagger \tilde{b}_v
\end{align}
where $\tilde{b}_v$ is the vortex of $\tilde{b}$. Now let us consider a FQH state, denoted $\mathcal{F}$, which can always be understood in terms of the above construction if $\tilde{b}$ forms some other topological phase of matter $\mathcal{B}$. The above suggests that the particle-hole conjugate of $\mathcal{F}$, which we denote as $\overline{\mathcal{F}}$, can be understood as a state where the vortices of $\tilde{b}$ form the bosonic state $\mathcal{B}$. In other words, $\overline{\mathcal{F}}$ is the resulting fermionic phase when $\tilde{b}$ forms the vortex dual of $\mathcal{B}$. 

Let us study several concrete examples. First, let us consider a generic Abelian fermionic FQH state described by a symmetric integer matrix $K$ and a charge vector $q$. This means that the effective theory of the electrons is
\begin{align}
\mathcal{L}_c = \frac{1}{4\pi} K_{IJ} a_I \partial a_J + \frac{1}{2\pi} q_I A^E \partial a_I, 
\end{align}
where $A_E$ is the external response field. In terms of our construction, this can be rewritten as
\begin{align}
\mathcal{L}_c = &-\frac{1}{4\pi} \tilde{a} \partial \tilde{a} + \frac{1}{2\pi} \tilde{A} \partial \tilde{a} + 
\nonumber \\
&\frac{1}{4\pi} K_{b;IJ} a_I \partial a_J + \frac{1}{2\pi} A^E \partial \tilde{a} + \frac{1}{2\pi} q_{b;I} \tilde{A} \partial a_I,
\end{align}
where $K_b$ is a symmetric even integer matrix and $q_b$ is an integer vector. The $U(1)$ gauge field $\tilde{a}$ describes the IQH state of $f$, while $\tilde{A}$ is the emergent gauge field associated with the gauge redundancy $f \rightarrow e^{i\theta} f$, $b \rightarrow e^{-i\theta} b$. Integrating out $\tilde{A}$ sets
$\tilde{a} = \sum_I q_{bI} a_I$, which then implies the relation
\begin{align}
K_{IJ} = K_{b;IJ} - q_{b;I} q_{b;J} ,
\end{align}
and $q = q_b$. 

Now, suppose that $b$ instead forms the vortex dual of the theory described by $(-K_b,- q)$. In this case, the electron theory would be
\begin{align}
\mathcal{L}_{ph} =  &-\frac{1}{4\pi} \tilde{a} \partial \tilde{a} + \frac{1}{2\pi} \tilde{A} \partial \tilde{a}  
-\frac{1}{4\pi} K_{b;IJ} a_I \partial a_J 
\nonumber \\
&+ \frac{1}{2\pi} A^E \partial \tilde{a} - \frac{1}{2\pi} q_{b;I} A \partial a_I
- \frac{1}{2\pi} \tilde{A} \partial A . 
\end{align}
Integrating out $\tilde{A}$ sets $A = \tilde{a}$; further integrating out $A$ then gives
\begin{align}
\mathcal{L} = \frac{1}{4\pi} (q_b^I q_b^J - K_{IJ}^b) a_I \partial a_J + 
\frac{1}{4\pi} A^E \partial A^E - \frac{1}{2\pi} q_{b;I} A^E \partial a_I, 
\end{align}
which is precisely the particle-hole conjugate of the theory described by $K$. 

The theory of the Anti-CFL developed in Sec. \ref{antiCFLbulk} provides another example of the general relation discussed here, 
where the Anti-CFL state can be understood as a state where the composite bosons $\tilde{b}$ form the vortex dual of a $\nu = -1$ bosonic CFL. 

\bibliography{TI}

\end{document}